**Popular individuals process the world in particularly normative ways**

Elisa C. Baek[1]*, Ryan Hyon[1], Karina López[1], Emily S. Finn[2], Mason A. Porter[3,4], and Carolyn Parkinson[1,5]*

[1]Department of Psychology, University of California, Los Angeles, [2]Department of Psychological and Brain Sciences, Dartmouth College, [3]Department of Mathematics, University of California, Los Angeles, [4]Sante Fe Institute, [5]Brain Research Institute, University of California, Los Angeles

* Corresponding authors

**Abstract**

People differ in how they attend to, interpret, and respond to their surroundings. Convergent processing of the world may be one factor that contributes to social connections between individuals. We used neuroimaging and network analysis to investigate whether the most central individuals in their communities (as measured by in-degree centrality, a notion of popularity) process the world in a particularly normative way. We found that more central individuals had exceptionally similar neural responses to their peers and especially to each other in brain regions that are associated with high-level interpretations and social cognition (e.g., in the default-mode network), whereas less-central individuals exhibited more idiosyncratic responses. Self-reported enjoyment of and interest in stimuli followed a similar pattern, but accounting for these data did not change our main results. These findings suggest that highly-central individuals process the world in exceptionally similar ways, whereas less-central individuals process the world in idiosyncratic ways.

# Introduction

Humans are incredibly social, and deficits in social connection have been linked to myriad negative consequences, including increased likelihood of morbidity and mortality[1–4]. Having many social ties (i.e., being popular[i], such as by having a high in-degree centrality in a social network by being nominated as a friend by many peers) is one factor that can protect against the detrimental consequences of social isolation and disconnection[5–9]. Differences in the extent of social connectedness occur in many human social networks[10–12], and it has been established that such differences are critical determinants for the well-being of individuals[5]. They can also have far-reaching consequences for the social networks in which individuals are embedded. For example, central individuals often have significant influence in shaping the opinions and attitudes of social groups[13–16].

Despite robust evidence for the benefits to one's health and well-being of being well-connected and the fact that well-connected individuals are well-positioned to exert influence on others in their social networks, there are significant gaps in our understanding of which factors distinguish popular individuals. For instance, although some personality traits (such as extraversion and emotional stability) have been associated with being well-connected in some social networks[17,18], such links have not been found in other contexts[19–21]. It is possible that approaches that focus on personality do not capture features that distinguish popular individuals across various social contexts. For example, one possibility is that individuals who occupy central[ii] positions in a social network process the world around them in a way that allows them to

---

[i] There are many ways of defining popularity. In the present paper, we use the notion of "popularity" as synonymous with having a high in-degree centrality. In our data, an individual's in-degree centrality is equal to the number of times that they were nominated as a friend by other community members. See the "Methods" section for more details.

[ii] In the present paper, we use the term "central" to refer to having a high in-degree centrality.

relate to, understand, and connect with a larger number of people in their communities. Recognizing and adhering to social norms is critical to being successful in forming and maintaining social ties[22], so popular individuals may be more attuned to their peers' norms either as a cause or as a consequence (or as a combination of both) of their central position in a network. Accordingly, popular individuals may process the world around them in ways that are exceptionally similar to their peers. Correspondingly, it is possible that less-popular individuals may process the world around them in ways that are less similar to their peers (including each other) than is the case for popular individuals. Therefore, less-popular individuals may hold less-central positions in their social networks because they process the world around them in a way that does not reflect of the norms of their peers (i.e., in a way that is more idiosyncratic than others).

In the present paper, we test the hypothesis that individuals who occupy central positions in their social networks have neural responses to naturalistic stimuli (specifically, videos) that are exceptionally similar to those of their peers (relative to individuals who occupy less-central positions). Specifically, we test whether individuals who many others nominate as a friend (i.e., who have a high in-degree centrality) have neural responses that are, on average, more similar to their peers than individuals who are unpopular in their social network (i.e., who fewer people indicate as a friend and thus have a low in-degree centrality). Measuring neural activity during a naturalistic paradigm (in which people view complex audiovisual stimuli, such as videos, that unfold over time) allows one to obtain insight into individuals' unconstrained thought processes as they unfold[23]. Coordinated brain activity between individuals (i.e., large inter-subject correlations (ISCs) of neural responses) during the viewing of dynamic, naturalistic stimuli has been associated both with friendship[24] and with shared interpretations and understanding of events[25–27]. Therefore, the extent to which an individual, on average, has similar neural-response time series as their peers can provide insight into the extent to which they process the world around them in a way that reflects the norms of their communities.

We also test whether individuals who are highly central in their social networks are exceptionally similar to other highly-central individuals in how they process the world around them whereas less-central individuals process the world around them in their own idiosyncratic ways. To help explain this idea, we draw an analogy from the famous opening line of the novel *Anna Karenina*: "Happy families are all alike; every unhappy family is unhappy in its own way.[28]" An Anna Karenina principle posits that endeavors with particular outcomes share similar characteristics (so, in that sense, they are "are all alike") and that a lack of any one of the characteristics results in the absence of the outcome in question[29]. The concept of an Anna Karenina principle has been applied to study various phenomena[30]. For example, it was used recently to link neural similarity with behavioral outcomes, such as trait paranoia[31]. In the present work, we test the hypothesis that "Popular individuals are all alike, but each unpopular individual is dissimilar in their own way."

We first test whether popular individuals in a community process the world around them in a way that is exceptionally similar to other community members. We assess this idea by calculating the mean neural similarity between them and their peers. (See "Subject-level ISC analysis" in the "Methods" section for more details.) We also test the hypotheses that popular individuals have exceptionally similar neural responses to each other and that each unpopular individual responds in their own unique way (i.e., idiosyncratically). Our results provide support for both hypotheses. We found that, on average, popular individuals had exceptionally similar neural responses to other members of their communities and especially to other popular individuals in brain regions that are associated with shared high-level interpretations and social cognition (e.g., regions of the default-mode network). By contrast, we found that less-central individuals had more idiosyncratic neural responses. We obtained similar results when we controlled for demographic similarities and social distances between individuals. Additionally, although participants' self-reported enjoyment of and interest in the stimuli followed a similar pattern as the brain data, accounting for their self-reported preferences did not change our main results. Taken together, our findings suggest that popular individuals tend to be exceptionally similar to each other in the ways that they process the world around them and that each unpopular individual is dissimilar in their own idiosyncratic way.

## Results

**Social-network characterization.** We characterized the social networks of individuals who live in two different residential communities of first-year students at a large state university (University of California, Los Angeles) in the United States. A total of 120 participants completed an online survey in which they indicated individuals with whom they were friends within their community. (See the "Methods" section for further details.) Some of these participants also completed the functional magnetic resonance imaging (fMRI) part of the study. (The fMRI part of the study included $N = 63$ people after exclusions; see the "Methods" section.) Using the responses of the participants, we constructed a directed network for each of the two communities (see Fig. 1). In each of these networks, a node represents an individual and a directed edge represents one individual nominating another as a friend. For each individual, we calculated in-degree centrality, which counts the number of times that the individual was nominated as a friend by someone else in the network. We chose to quantify an individual's popularity within their community in terms of in-degree centrality because it captures the extent to which others in the community consider the individual to be a friend. Another advantage of in-degree centrality is that an individual's in-degree centrality (unlike some other measures of centrality, such as out-degree centrality) does not rely at all on one's own self-reported answers about the relationships that one has with others. Therefore, in-degree centrality is not susceptible to erroneous perceptions of one's own friendships and is less susceptible to the mischaracterization of friendship ties due, for example, to any given participant's inattention during a survey or atypical interpretations of survey questions (because an individual's in-degree centrality is based on data that is aggregated across the responses of many participants). Additionally, in-degree centrality is particularly suitable for our study because it is not affected by the presence of multiple components in a network, unlike most other measures of centrality (e.g., eigenvector centrality)[32].

In our primary analyses, we used a median split to binarize our sample into high-centrality and low-centrality groups. This choice is consistent with recent studies that related neural similarity with behavioral measures[33,34]. In our fMRI study, we classified participants as part of the high-centrality group if they had an in-degree that was larger than the median (specifically, if it was more than 2; there were $n_{high} = 23$ such people) and into the low-centrality group if they had an in-degree that was less than or equal to the median (specifically, if it was less than or equal to 2; there were $n_{low} = 40$ such people). See Supplementary Fig. 1 for plots of the in-degree distributions. Because the median-split approach resulted in unevenly-sized groups, we also conducted additional analyses to examine the relationships between the original non-binarized version of centrality and neural similarity whenever possible, as we describe in more detail below. We also conducted analogous exploratory analyses with approximately equal-sized groups by contrasting individuals with in-degree centralities in the top and bottom thirds of the distribution. This yielded similar results to our main findings. See "Subject-level ISC analysis" and "Dyad-level ISC analysis" for more details.

**Neural similarity.** During our fMRI study, participants watched 14 video clips that span a variety of topics (see Supplementary Table 1). We calculated inter-subject correlations (ISCs) of time series of neural responses that were measured with fMRI to capture shared neural responses across subjects during the processing of naturalistic stimuli[35] (see Fig. 1). First, we extracted the mean-response time series across the entire video-viewing task from both (1) each of the 200 cortical regions in the 200-parcel version of the Schaefer et al. (2018)[36] parcellation scheme and (2) 14 subcortical regions[37]. (See the "Methods" section for more details.) This resulted in a total of 214 brain regions across the whole brain. For each of the 1,952 unique pairs of participants (i.e., dyads[iii]) in our fMRI sample, we then computed the Pearson correlation between the dyad members' time series of neural responses for each cortical region. This yields one correlation coefficient per unique dyad for each brain region. See the "Methods" section for more details.

---

[iii] The term "dyad" is sometimes used to refer specifically to an adjacent pair of nodes in a network (i.e., to include both nodes and the edge that connects them). One can think of the set of ISCs between all participants for a given brain region as a complete weighted graph in which edge weights encode ISCs. Therefore, we refer to each possible pair of fMRI participants (whether or not there was a friendship connection between them) as a "dyad".

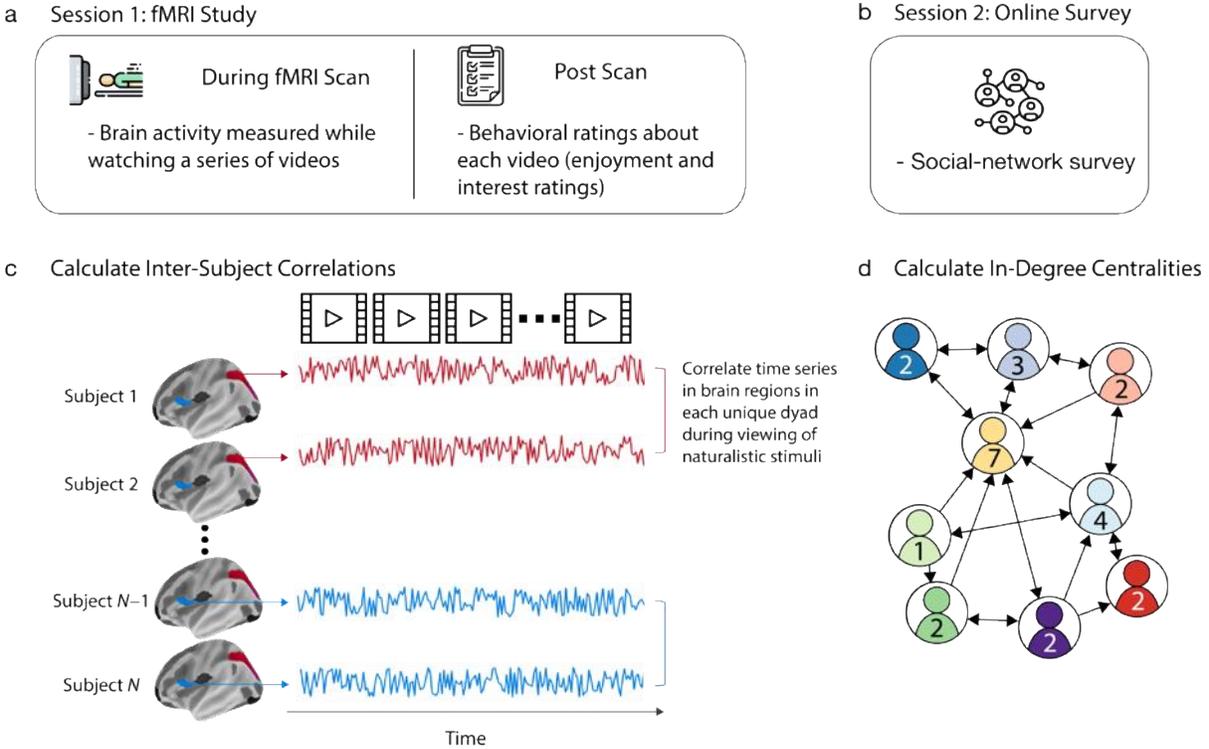

**Fig. 1. Study paradigm and calculations. (a)** Schematic of the fMRI study paradigm. In session 1 of the study, participants attended an in-lab session in which their brain activity was measured using fMRI while they watched a series of naturalistic stimuli (i.e., videos). After the fMRI scan, the participants provided ratings on how enjoyable and interesting they found each video. **(b)** Schematic of our social-network survey. In session 2 of the study, participants completed an online social-network survey in which they indicated the individuals in their residential community with whom they were friends. **(c)** Schematic of neural similarity. We extracted the time series of neural responses that were obtained as participants viewed the stimuli. We then calculated inter-subject correlations (ISCs) of these time series for each of 214 brain regions. **(d)** Schematic of our network calculations. Based on the participants' responses in **(b)**, we constructed two directed, unweighted networks — with one for each residential community — in which each node represents an individual and each directed edge represents one individual nominating another as a friend. For each individual, we calculated in-degree centrality, which counts the number of times that that individual was nominated as a friend by others in their own residential community.

**Subject-level ISC analysis.** We tested whether individuals who were more popular in their communities (i.e., who had higher in-degree centralities) exhibited more normative neural responses than less-popular individuals (i.e., those with lower in-degree centralities). To do this, in each brain region, we transformed our dyad-level neural similarity measure to a subject-level measure by calculating the mean Fisher *z*-transformed[38] ISC value for each subject with each other subject.. This yields one ISC value for each individual for each brain region; this value encodes a mean similarity in neural responses between the individual and all other individuals in the corresponding brain region (see Fig. 2a). After calculating these values, we fit one generalized linear model (GLM) for each brain region with the ISC in the respective brain region as the dependent variable (which we transformed into z-scores to produce standardized coefficients) and the binarized in-degree as the independent variable (see Fig. 2b). Finally, we employed false-discovery rate (FDR) correction to correct for multiple comparisons across brain regions. We found that high in-degree centrality was associated with larger mean neural similarity with peers in the dorsomedial prefrontal cortex (DMPFC) bilaterally (left DMPFC: $B = 0.964$, $SE = 0.233$, $p_{corrected} = 0.012$, $p_{uncorrected} < 0.001$; right DMPFC: $B = 0.977$, $SE = 0.232$, $p_{corrected} = 0.012$, $p_{uncorrected} < 0.001$) and right precuneus ($B = 0.912$, $SE = 0.237$, $p_{corrected} = 0.020$, $p_{uncorrected} < 0.001$) (see Fig. 2c). We did not find any significant associations in the subcortical regions (see Supplementary Table 2). We also fit analogous models to control for demographic variables that may be associated with neural similarity[24,39], models that only incorporated neural similarities between subjects who were living in the same residential community, models that controlled for social distances between participants in the same community, and models that used a subset of the data with approximately equal-sized centrality groups. These other approaches yielded similar results. See Supplementary Figs. 2–5.)

To confirm that our results from analyzing binarized-centrality groups also hold when we treat in-degree centrality in its original (i.e., non-binarized) form, we also conducted an analogous analysis to relate participants' mean ISCs with each other in each brain region with the non-binarized in-degree centrality values. For each brain region, we calculated the Spearman rank correlation ρ to examine the relationship between ISCs in each brain region and in-degree centrality. We again employed FDR correction to correct for multiple comparisons across brain regions. Using these computations, we identified similar regions as when we used binarized in-degree centrality (i.e., as low versus high values). We found that neural similarities in the bilateral DMPFC (left DMPFC: $\rho = 0.420$, $p_{corrected} = 0.048$, $p_{uncorrected} < 0.001$; right DMPFC: $\rho = 0.415$, $p_{corrected} = 0.048$, $p_{uncorrected} < 0.001$), precuneus ($\rho = 0.408$, $p_{corrected} = 0.048$, $p_{uncorrected} < 0.001$), and left superior parietal lobule ($\rho = 0.424$, $p_{corrected} = 0.048$, $p_{uncorrected} = 0.002$) were significantly correlated with in-degree centrality (see Fig. 2d). In other words, we found that there was a positive association between an individual's in-degree centrality and their mean neural similarity with their peers in the DMPFC, precuneus, and superior parietal lobule. See Fig. 3 for a visualization of the ISC in the right DMPFC and its association with in-degree centrality. We did not find any significant associations in subcortical regions (see Supplementary Table 3).

Notably, for both sets of analyses, we found a positive relationship in all cases in which participants' ISCs with their peers were related significantly to their in-degree centrality. That is, in both analyses, we found that a higher in-degree centrality was associated with more normative neural responses. Additionally, in these analyses and in all of our other analyses, we did not find any regions in the brain in which low in-degree centrality was associated with more-similar neural responses to one's peers.

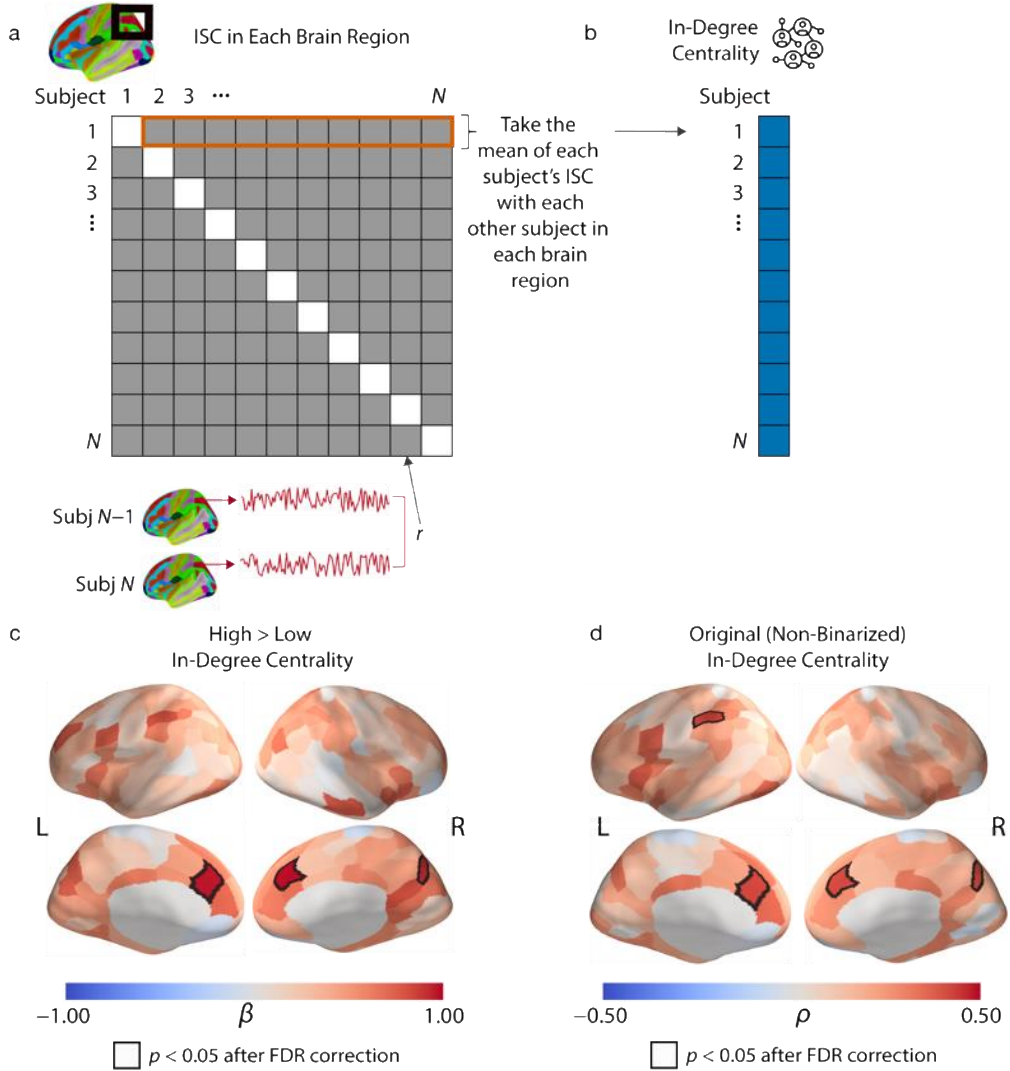

**Fig. 2. Subject-level analysis**. **(a)** Our approach for subject-level analysis. First, we Fisher z-transformed the dyad-level inter-subject correlations, which are encoded by a matrix of pairwise Pearson correlation coefficients (which we denote by *r*). We then computed the mean of each subject's ISC with each other subject. (In other words, we took the mean of each row of the matrix.) We performed the above calculations for each of the 214 brain regions. This yields one ISC value for each subject for each brain region. The ISC value encodes the mean similarity in neural responses between the subject and each other subject in the corresponding brain region. **(b)** We tested for relationships between the subjects' in-degree centrality and these subject-level ISC values in each brain region. **(c)** Our results that relate mean ISCs with the binarized in-degree centrality variable indicated that individuals with high in-degree centrality had a much larger mean neural similarity with their peers in the bilateral DMPFC and precuneus than individuals with a low in-degree centrality. **(d)** Our results that relate mean ISCs with the original (i.e., non-binarized) in-degree centrality values gave similar results as the analysis in **(c)**. We found that the mean ISCs in the bilateral DMPFC, precuneus, and the superior parietal lobule were positively correlated with in-degree centrality. The quantity *B* denotes the standardized regression coefficient, and ρ denotes the Spearman rank correlation. All results are FDR-corrected at $p < 0.05$, which corresponds to an uncorrected *p*-value of 0.009 in (c) and an uncorrected *p*-value of 0.001 in (d).

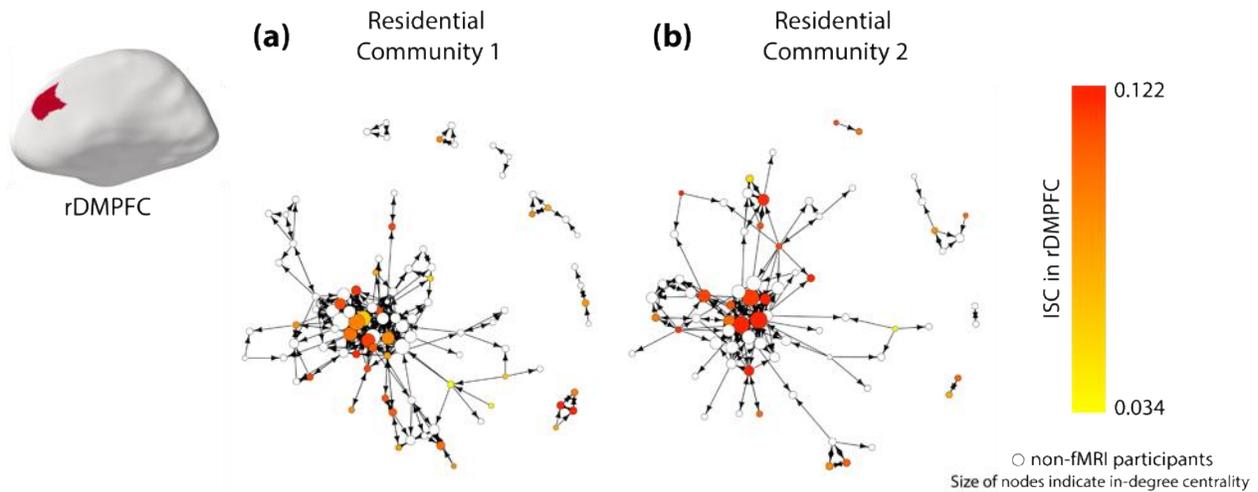

**Fig. 3. Visualization of subject-level ISC results in the social networks.** Visualizations of the social networks of **(a)** residential community 1 and **(b)** residential community 2 of a first-year dorm. Each subject was a resident of one of two distinct residential communities, where one "community" consists of the set of people who live in the same wing and floor of a residence hall. Each node (which we show as a disc) represents one resident who was living in one of the communities, and each line segment represents one directed edge between two nodes if it is unidirectional and represents two directed edges if it is bidirectional. For example, an arrow from node A to node B conveys that node A nominated node B as a friend. An edge with two arrowheads indicates a mutually-nominated friendship. The size of a node represents its in-degree centrality, with larger nodes indicating individuals with higher in-degree centrality. The color of the nodes represents a node's mean neural similarity in the rDMPFC to other members of its residential community, with darker colors indicating greater neural similarity. As this figure indicates, individuals who had higher in-degree centrality (i.e., individuals who many other individuals nominated as a friend) tended to have the largest mean ISCs with their peers in the rDMPFC.

**Preference similarity.** After the neuroimaging portion of the fMRI study, participants rated the extent to which they felt that each video that they saw in the scanner was enjoyable and interesting. For each subject, we took the following steps to calculate their mean similarity with their peers in enjoyment and interest ratings. For each of the 1,952 unique dyads (i.e., pairs of individuals), we calculated the Euclidean distance between the two subjects' enjoyment ratings across the 14 different videos and transformed the distance measure into a normalized similarity measure (where the similarity is $s = 1 - [\text{distance} / \max(\text{distance})]$). Larger similarity values, which range from 0 to 1, indicate greater similarity in how much two individuals in a dyad enjoyed the content. We repeated the same process for interest ratings. This yields two preference similarity measures per dyad.

**Analysis of subject-level preferences.** We were interested in (1) whether individuals who were more popular in their residential community had preferences that were more similar to others in their community than less-central individuals and (2) if such self-reported differences in preferences could account for the neural results that we reported above. To investigate this, we transformed the dyad-level preference similarity measures to subject-level variables. First, for each subject, we calculated their mean similarity in enjoyment ratings with each other subject. This estimates the extent to which that subject, on average, had similar preferences to other subjects in how enjoyable they found the videos. We repeated the same process for the interest ratings. This approach yields one number for each subject to represent their mean similarity with their peers in enjoyment ratings and one number to represent their mean similarity with their peers in interest ratings. We then related the mean enjoyment and interest similarity measures with the binarized in-degree centrality variable by fitting a GLM for each similarity measure with z-scores of the similarity measures as the dependent variables and the binarized in-degree centrality variable as the independent variable. Our results indicate that individuals who were more popular in their social networks were more similar, on average, than less-popular individuals with their peers in the content that they found to be enjoyable ($B = 0.578$, SE = 0.253, $p = 0.026$) and interesting ($B = 0.491$, SE = 0.256, $p = 0.061$). Note that the association between popularity and mean interest similarity is only marginally statistically significant (i.e., trend-level.)

Given our finding that individuals with a high in-degree centrality were more similar to their peers in self-reported content preferences than those with a low in-degree centrality, we tested whether our findings that link ISC to in-degree centrality could arise from inter-subject similarities in self-reported preferences. To investigate this possibility, we fit GLMs to test the relationship between the ISC in each brain region and in-degree centrality while controlling for similarity in enjoyment and interest ratings. Our results indicate that the relationships between ISC and in-degree centrality remain significant after controlling for similarity in enjoyment and interest ratings (see Supplementary Fig. 6), suggesting that neural similarity in these regions captures similarities beyond what one can attribute purely to self-reported preference ratings.

**Dyad-level ISC analysis.** Our subject-level ISC results indicate that individuals with a higher in-degree centrality (i.e., those who were nominated as a friend by many individuals) had, on average, greater neural similarity with their peers than individuals with a lower in-degree centrality. We also took a finer-grained approach to test if individuals with similar in-degree centralities were most similar to one another, irrespective of whether they had a high or a low in-degree centrality, or if individuals who were highly central in their residential community were most similar to other highly-central individuals and less-central individuals were comparatively idiosyncratic (i.e., dissimilar to others, including other individuals with low in-degree centralities). To relate our dyad-level neural similarity measure with individuals' in-degree centralities, we transformed the subject-level binarized in-degree centrality measure into a dyad-level variable. We categorized the dyads into (1) {high, high} if both subjects in the dyad had a high in-degree centrality, (2) {low, low} if both subjects in the dyad had a low in-degree centrality, and (3) {low, high} if one subject in the dyad had a low in-degree centrality and the other subject had a high in-degree centrality. For each of our 214 brain regions, we fit a linear mixed-effects model with crossed random effects to account for the dependency structure of the data[40] (see the "Methods" section) with ISC in the corresponding brain region as the dependent variable and the dyad-level centrality variable as the independent variable. We then performed a planned-contrast analysis[41] to compare the different in-degree centrality groups and thereby identify brain regions for which including one or more low-centrality individuals in a dyad was associated with less-coordinated neural responses (i.e., $ISC_{\{high, high\}} > ISC_{\{low, low\}}$, $ISC_{\{high, high\}} > ISC_{\{low, high\}}$, and $ISC_{\{low, high\}} > ISC_{\{low, low\}}$) (see Fig. 4a).

The $ISC_{\{high, high\}} > ISC_{\{low, low\}}$ contrast is our most direct test of the hypotheses that highly-central individuals have exceptionally similar neural responses to one another, whereas less-central individuals have neural responses that are idiosyncratic. This is the case because it tests whether neural similarity is greater in dyads in which both individuals had a high in-degree centrality than in dyads in which both individuals had a low in-degree centrality. By contrast, $ISC_{\{high, high\}} > ISC_{\{low, high\}}$ would also hold for a nearest-neighbor model[31], which reflects the assumption that individuals who are more similar in a behavioral trait also exhibit greater neural similarity[iv], and $ISC_{\{low, high\}} > ISC_{\{low, low\}}$ does not necessarily have to hold to support the hypotheses that highly-central individuals have exceptionally similar neural responses to one another but that less-central individuals have neural responses that are idiosyncratic. For example, if each low-centrality subject responded in a completely unique way, then they would have similarly low ISCs with other low-centrality individuals and with high-centrality individuals). With that said, we reasoned that $ISC_{\{low, high\}} > ISC_{\{low, low\}}$ was likely to arise in the current dataset because of underlying stimulus-driven responses that are shared across all participants, and each low-centrality individual will partially reflect these shared stimulus-driven responses (even if they each deviate from the normative responses in an idiosyncratic way). Accordingly, we report the results of three contrasts: (1) $ISC_{\{high, high\}} > ISC_{\{low, low\}}$, which is the most direct test of our hypotheses; (2) $ISC_{\{high, high\}} > ISC_{\{high, low\}}$, which is a test of our

---

[iv] We use the term "nearest-neighbor model" to refer to the assumption that individuals should always be most similar to their immediate neighbors in terms of their neural responses, regardless of their absolute position on some scale (e.g., whether they are high or low in in-degree centrality). See [31] for further discussion.

hypotheses but also holds for a nearest-neighbor model; and (3) $ISC_{\{low, high\}} > ISC_{\{low, low\}}$, which does not have to hold to support our hypotheses, but which we expect to hold.

We illustrate the results of the three contrasts ($ISC_{\{high, high\}} > ISC_{\{low, low\}}$, $ISC_{\{high, high\}} > ISC_{\{low, high\}}$, and $ISC_{\{low, high\}} > ISC_{\{low, low\}}$) in Fig. 4. As in our subject-level results, our dyad-level results reveal that there were larger ISCs in the DMPFC, precuneus, and portions of the superior parietal lobule in dyads of individuals who both had high in-degree centralities (i.e., {high, high}) than in dyads of individuals who both had low in-degree centralities (i.e., {low, low}) (see Fig. 4b). Additionally, ISCs in the ventrolateral prefrontal cortex (VLPFC) and temporal pole were larger in {high, high} dyads than in {low, low} dyads. ISCs in subcortical regions (including the amygdala, hippocampus, left pallidum, and the right thalamus) were larger in {high, high} dyads than in {low, low} dyads (see Supplementary Table 4). We found similar patterns when we contrasted high-centrality dyads with mixed-centrality dyads ($ISC_{\{high, high\}} > ISC_{\{low, high\}}$) and mixed-centrality dyads with low-centrality dyads ($ISC_{\{low, high\}} > ISC_{\{low, low\}}$), although the effect sizes were smaller. See Figs. 4b,c and Supplementary Tables 5–6. In the Supplementary Information, we report results of analogous models that control for demographic variables and friendship (see Supplementary Fig. 7) and that examine neural similarities only in subjects who live in the same residential community (see Supplementary Fig. 8). The latter approached allowed us to control for both demographic similarities and social distances between individuals (see Supplementary Fig. 9). We also report the results of models that use a subset of the data with approximately equal-sized centrality groups (see Supplementary Fig. 10). The results of these additional analyses are similar to those in Fig. 4. Across all of our analyses, we did not find any regions in the brain in which there were larger ISCs in {low, low} dyads than in {high, high} dyads. We also did not find any regions in the brain in which there were larger ISCs in {low, high} dyads than in {high, high} dyads, nor any in which there were larger ISCs in {low, low} dyads than in {low, high} dyads. Our findings suggest that highly-central individuals were exceptionally similar in neural responses to one another, whereas less-central individuals had neural responses that were dissimilar both to highly-central individuals and to other less-central individuals. In other words, less-central individuals had neural responses that were idiosyncratic, with each less-central individual differing from the normative response of other individuals in their own way.

We also conducted an analogous analysis to relate mean ISC with the original non-binarized values of the dyad-level in-degree centralities. To do this, we related the minimum in-degree centrality of each dyad to neural similarity in each of our 214 brain regions. Taking the minimum in-degree centrality value of each dyad allowed us to test the hypothesis that only dyads with two highly-central individuals had exceptionally similar neural responses to one another. If a low in-degree centrality is associated with an idiosyncratic neural response, then the inclusion of even just one low-centrality individual in a dyad should be associated with a small ISC. For each brain region, we fit a linear mixed-effects model with crossed random effects to account for the dependency structure of the data[40] (see the "Methods" section) with the ISC in the corresponding brain region as the dependent variable and the log-transformed[v] minimum in-degree centrality value of each dyad as the independent variable.

As with our dyad-level results using the binarized centrality variable, we found a positive association between the minimum in-degree centrality of the dyads and neural similarity in the left DMPFC, precuneus, posterior cingulate cortex, superior parietal lobule, and middle temporal gyrus. That is, there was greater neural similarity in these brain regions in dyads with a higher minimum in-degree centrality. Mirroring our results with a binarized in-degree centrality variable, dyads in which both individuals were highly central in their residential community (as encoded by a higher minimum in-degree centrality) had greater neural similarity than dyads in which both individuals were less central (as encoded by a lower minimum in-degree centrality) (see Fig. 5).

---

[v] Specifically, we used $\ln(1 + \text{in-degree centrality})$.

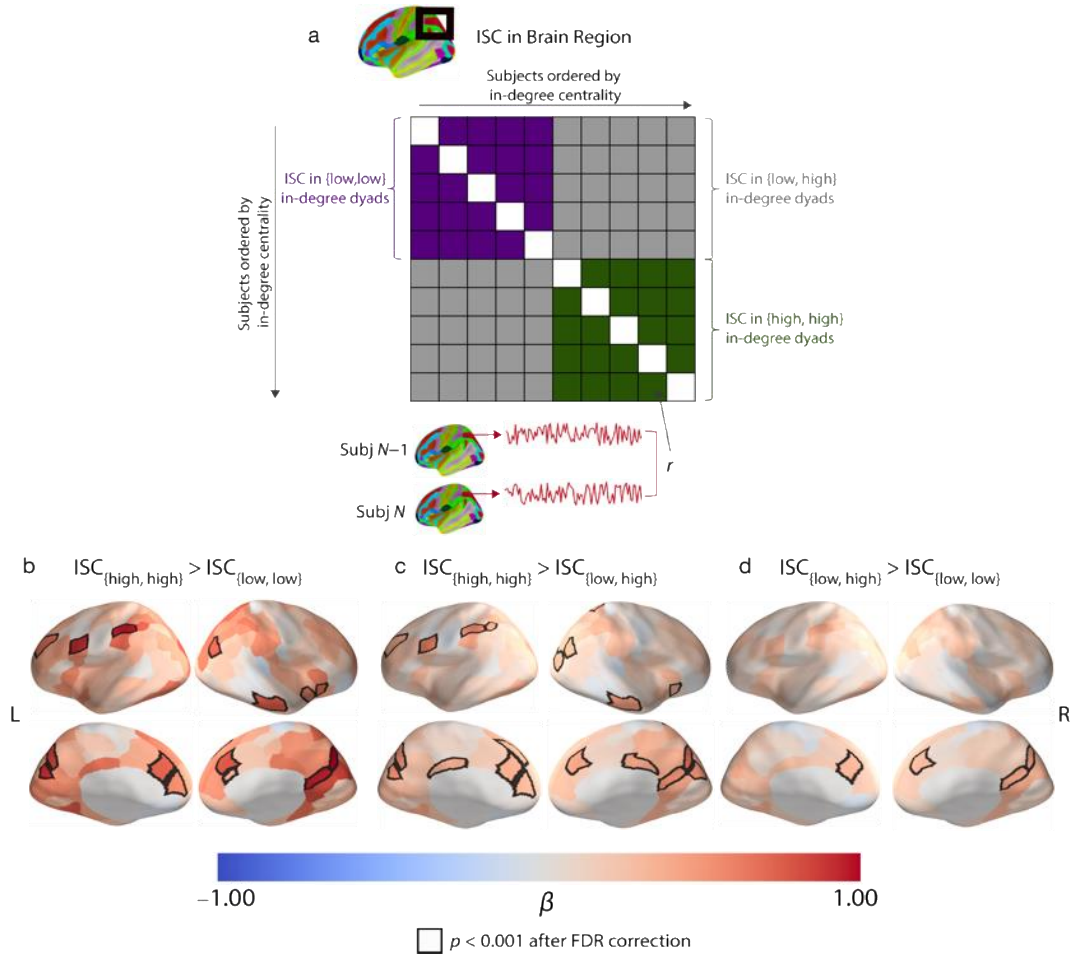

**Fig. 4.** Dyad-level analysis and results. (a) Dyad-level ISCs in a brain region are encoded in a matrix whose entries consist of pairwise Pearson correlation coefficients. The rows and columns of the matrix are ordered according to the in-degree centralities of the subjects. We performed planned contrasts of the different centrality groups to test whether dyads in which both individuals were highly central (i.e., $ISC_{\{high, high\}}$), had larger ISCs than dyads in which both individuals were less central (i.e., $ISC_{\{low, low\}}$) and than dyads with mixed centralities (i.e., $ISC_{\{low, high\}}$), for which one individual of the dyad had a low centrality and the other had a high centrality. [The figure in (a) is adapted from prior work[33].] (b) There were larger ISCs in the DMPFC, VMPFC, VLPFC, precuneus, temporal pole, and portions of the superior parietal lobule in {high, high} dyads than in {low, low} dyads. (c) We found similar patterns when we compared {high, high} dyads to {low, high} dyads and (d) when we compared {low, high} dyads to {low, low} dyads. The $ISC_{\{high, high\}} > ISC_{\{low, low\}}$ contrast in (b) provides the most direct test of our main hypotheses that highly-central individuals have exceptionally similar neural responses to one another and that less-central individuals have neural responses that are idiosyncratic.

The quantity $B$ is the standardized regression coefficient. Regions with significant differences for each contrast are outlined in black. We used an FDR-corrected significance threshold of $p < 0.001$, which corresponds to an uncorrected $p$-value threshold of $p < 6.386 \times 10^{-5}$.

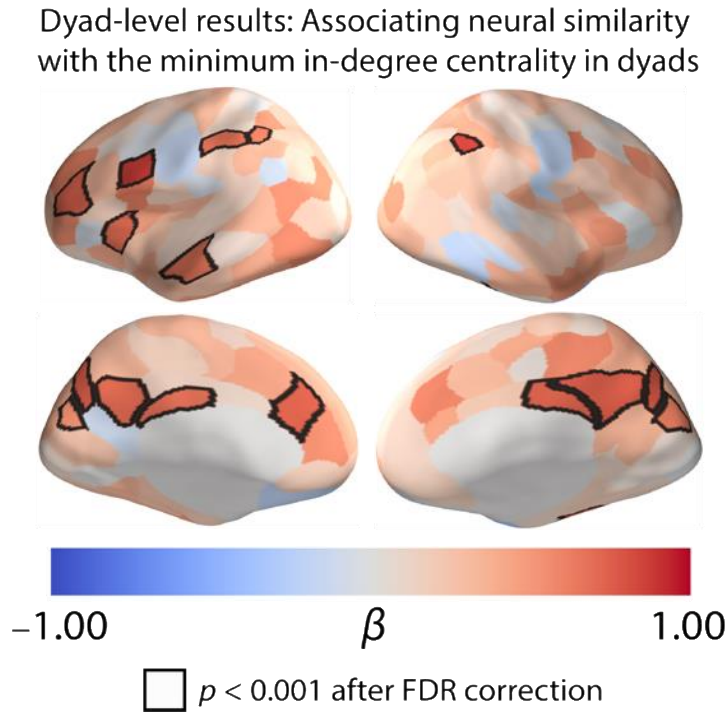

**Fig. 5.** Dyad-level associations of neural similarity with the minimum in-degree centrality of dyads. We found a positive association between ISC and minimum in-degree centrality. Larger ISCs in brain regions (including the DMPFC, the VLPFC, the precuneus, the temporal pole, and portions of the superior parietal lobule) were associated with a higher minimum in-degree centrality. The quantity $B$ is the standardized regression coefficient. Regions where we observed significant associations between in-degree centrality and ISC are outlined in black. We used an FDR-corrected significance threshold of $p < 0.001$, which corresponds to an uncorrected $p$-value threshold of $p < 8.879 \times 10^{-5}$.

**Analysis of dyad-level preferences.** We tested whether the self-reported preferences of individuals were consistent with the hypotheses that more-central individuals have preferences that are exceptionally similar to each other and that less-central individuals have preferences that are idiosyncratic, with each low-centrality individual's preferences differing from those of other individuals in their own way. We also tested if such self-reported differences in preferences could account for our neural results. We first fit two mixed-effects models, with crossed random effects to account for the dependency structure of the data[40]. (See the "Methods" section.) We employed one such model for each type of preference (i.e., similarities in enjoyment and interest ratings). We used dyad-level similarities in enjoyment and interest ratings (see the above discussion of "Preference similarity") as the dependent variables — one in each of the two models — and the dyad-level minimum-centrality variable as the independent variable. We then performed planned contrasts of the three different dyad-level centrality groups (i.e., {high, high}, {low, low}, and {low, high}) to test if the inclusion of one or more low-centrality individuals in a dyad was associated with lower levels of interpersonal similarities in preferences (i.e., $s_{\{high, high\}} > s_{\{low, low\}}$, $s_{\{high, high\}} > s_{\{low, high\}}$, and $s_{\{low, high\}} > s_{\{low, low\}}$, where $s$ corresponds to dyad-level preference similarity, as defined in the above discussion of "Preference similarity"). We employed FDR correction to correct for multiple comparisons from the multiple planned contrasts. Our results indicate that dyads that consisted of two highly-central individuals (i.e., {high, high}) were more similar to one another in what they found enjoyable and interesting than dyads that consisted of two less-central individuals ({low, low}) (see Supplementary Tables 7 and 8). We found similar patterns when we compared high-centrality dyads to mixed-centrality dyads and when we compared mixed-centrality dyads to low-centrality dyads (see Supplementary Tables 7 and 8).

We then tested whether the above behavioral findings, which (like our neural findings) support the hypotheses that highly-central individuals were exceptionally similar in their preferences and that less-central individuals had idiosyncratic preferences, could account for our neural results. Specifically, we examined whether inter-subject similarities in self-reported preferences could explain our observation that individuals who were highly central in their residential community were exceptionally similar in their neural responses to other highly-central individuals and that less-central individuals were comparatively idiosyncratic. To examine this possibility, we fit additional linear mixed-effects models to test the relationship between ISCs in each brain region and dyad-level in-degree centrality (i.e., whether a given dyad was composed of two high-centrality individuals, two low-centrality individuals, or one high-centrality individual and one low-centrality individual) while controlling for similarity in enjoyment and interest ratings. Although similarity in enjoyment and interest ratings were also associated with neural similarity in regions of the default-mode network (see Supplementary Figs. 10 and 11), our results indicate that the "Anna Karenina" pattern of results that links ISCs and dyad-level in-degree centralities remains significant after controlling for similarity in enjoyment and interest ratings (see Supplementary Fig. 13). Therefore, we conclude that our findings that greater neural similarity tends to occur between highly-central individuals and that reduced neural similarity tends to occur between less-central individuals arise from differences beyond those that were captured by self-reported preference ratings.

**Discussion**

What factors distinguish highly-central individuals in social networks? Our results are consistent with the notion that popular individuals (who are central in their social networks) process the world around them in normative ways, whereas unpopular individuals process the world around them idiosyncratically. In our study, we found that popular individuals exhibited greater mean neural similarity with their peers than unpopular individuals in several regions of the brain, including ones in which similar neural responding has been associated with shared high-level interpretations of events and social cognition (e.g., regions of the default-mode network) while viewing dynamic, naturalistic stimuli[42]. We observed a distinct pattern in the relationship between popularity and neural similarity: popular individuals were very similar to each other in their neural responses, whereas unpopular individuals were dissimilar both to each other and to their peers' normative way of processing the world. Our findings are consistent with the possibility that highly-central people process and respond to the world around them in a manner that allows them to relate to and connect with many of their peers and with the possibility that less-central people have idiosyncrasies that may result in greater difficulty in relating to others.

Brain areas in which popular individuals exhibited, on average, greater neural similarity with their peers than was the case for unpopular individuals included the bilateral DMPFC and precuneus, which are both regions of the default-mode network. Mirroring our findings that link popularity and mean neural similarity with community members, brain areas in which highly-central individuals responded exceptionally similarly to each other and less-central individuals responded idiosyncratically include the DMPFC, precuneus, and other regions of the default-mode network (such as the posterior cingulate cortex and the inferior parietal lobule). These regions have been implicated in social cognitive processes such as mentalizing and perspective-taking[43–45]. Neural similarity in these regions has also been associated with similarities in the understanding and interpretation of narratives, presumably because people who share similar viewpoints and perspectives have greater similarity in these higher-order brain regions during the viewing of naturalistic stimuli than those who do not[27,33,46]. Additionally, neural similarity in these regions has been associated with friendship[24]; friends appears to have greater similarity in these regions than people who are not friends. Of particular relevance to the present study, it was suggested recently[42] that the default-mode network helps promote a critical "sense-making" function by combining external information about one's surroundings with internal experiences and schemas to create models of situations as they unfold over time and that ISCs in such regions support the creation of shared meaning across individuals. Our results suggest that popular individuals process the world around them in a way that closely reflects their peers' normative way of understanding and responding to the world. Such similarity may help them relate and connect to many people (although further work is needed to elucidate the causal mechanisms that underlie our results). Our findings also suggest that popular individuals are exceptionally similar to each other and that less-popular individuals are dissimilar from a group's normative ways of processing and understanding the world (such that they process and respond to the world around them in their own idiosyncratic way). Our results were significant even when we controlled for (1) demographic variables that may be associated with neural similarity and (2) social distances between individuals. Therefore, our findings suggest that the association of neural similarity in regions of the default-mode network (and in other regions) with popularity is not merely a result of the most popular individuals being more likely to be friends with one another. Instead, we observed that highly-central individuals had exceptionally similar neural responses to those of many of their peers, including those with whom they were not friends.

In our study, popular individuals also self-reported preferences for the stimuli that were more reflective of the norms of their peers. Specifically, popular individuals had greater mean similarity with their peers in the extent to which they found stimuli to be enjoyable and interesting. Furthermore, popular individuals had exceptionally similar preferences for the stimuli as one another, but each unpopular individual had idiosyncratic preferences for the stimuli (i.e., preferences that were different both from the preferences of their peers on average and from those of other unpopular individuals). In concert, the observed behavioral patterns suggest that highly-central individuals self-report preferences that are more aligned with their peers' preferences and thus more "in tune" with what others find enjoyable or interesting; this may help them connect with their peers through mutually shared interests. Notably, controlling for similarities in the enjoyment and interest ratings did not change our results that link neural similarity with popularity. That is, we found that neural similarity in brain regions that have been implicated in high-level interpretation and social cognition was associated with network centrality above and beyond what we were able to capture using self-reported preferences. This suggests that measuring neural responses to naturalistic stimuli as they unfold over time can capture consequential aspects of mental processing beyond what one can obtain using a few targeted self-report questions. The strong link between popularity and ISCs (even when controlling for similarities in participants' self-reported preferences), relative to links between similarities in popularity and self-reported preferences, may arise from several factors, including the finer temporal granularity of ISCs than our self-report measures (because ISCs capture similarities in how responses evolve over time), the limits of self-report (because people are often unaware of and/or unwilling to report features of their attitudes and other aspects of their mental processing[47]), and the possibility that the similarities in processing that are linked to centrality reflect similarities in the creation of internal models of situations as they evolve over time (rather than reflecting similarities in what participants found interesting or enjoyable)[42]. A notable benefit calculating ISCs is that one can use them to characterize similarities in many different aspects of mental processing in parallel, and one can thereby obtain insight into diverse emotional and cognitive processes that unfold in response to various situations and which may be shaped by individuals' pre-existing beliefs, values, attitudes, and experiences. Our neural findings also provide insights that can inform which self-report measures are likely to capture the types of processing that may be particularly similar among popular individuals. It may be particularly fruitful to test for associations between popularity and the normativity of (1) self-report scales that capture individuals' social and cognitive tendencies and/or (2) individuals' processing of stimuli (e.g., as captured by semantic analyses of free-response measures).

In the present study, we obtained data from two different residential communities and characterized the neural similarity of each participant in our study with each other participant, including individuals from the other community. Furthermore, we successfully replicated all of our main effects that link ISCs and popularity when we fit models using only intra-community dyads. (See Supplementary Figs. 3 and 8.) The two residential communities each consisted of first-year students who were attending the same university. Although each residential community is relatively bounded and interactions between community members were likely to be uncommon — both because of restrictions that arose from the building structure and because of programming that focused on intra-community social activities — it is likely that the two residential communities had similar norms. Therefore, the types of normative processing that are associated with popularity in one residential community are likely to be similar to those that are associated with popularity in the other community. However, in some contexts, looking specifically at only intra-community similarities in neural activity may be important when relating ISCs with popularity, particularly when drawing on participants from communities with norms that are markedly different from each other. Future work can further elucidate the extent to which ISCs within and between communities are associated with individual differences in the centralities of individuals in social networks.

Does processing the world particularly normatively cause certain individuals to become highly central in their social networks, does being highly central in a social network cause certain individuals to process the world more normatively, or is it some combination of the two? Moreover, if popularity causally increases normative processing, do popular people (as a result of their central positions in a network) exert influence on other community members so that many individuals in the network become more similar to the popular people, do popular people change the way that they process the world around them to fit the norms of a social network, or is it some combination of the two?

Because the present study has only one wave of fMRI data, we are not able to ascertain the causal mechanisms that drive our effects. Prior research suggests that popular individuals have more behavioral and neural sensitivity than unpopular people to interpersonal cues[48] and that highly-central individuals are more likely than less-central individuals to adapt their brain activity to match that of other individuals in their social group[49]. Therefore, one possibility is that people who become popular may adapt their views of the world to meet their social network's normative ways of processing the world, perhaps due to a greater need to belong socially or a desire to connect with a large number of people. Future studies that employ longitudinal data can help elucidate the direction(s) of these effects and further clarify the mechanisms that may be at play. For instance, pairing in-lab measures of social conformity with longitudinal studies that examine neural similarity may provide insight into the extent to which individual differences in in-lab measures of conformity are predictive of changes in peoples' unconstrained processing of the world around them. Another possibility is that popular individuals may show similarly high levels of social abilities and functioning, which may in turn lead to greater neural similarity. For instance, it is possible that popular individuals may have distinctively high levels of empathic concern, mentalizing abilities, and/or emotion-regulation abilities that help them form and maintain a large number of social ties and also impact how they respond to naturalistic stimuli. Therefore, potential differences in social functioning between popular and other individuals may be a key reason for the links between popularity and neural similarity that we found in the present study. We did not collect measures that can capture individual differences in social functioning, so we are not able to test these theories using our data. Future studies that investigate associations between individual differences in social functioning, centrality, and neural processing of naturalistic stimuli can further elucidate these relationships.

In summary, our results suggest that highly-central individuals in a social network are exceptionally similar to their peers in how they process the world, as indicated by neural responses to naturalistic stimuli in brain regions that are associated with social cognition and building shared internal models of situations. We also found support for the idea that highly-central individuals are exceptionally alike in their neural processing and that less-central individuals are each dissimilar from their social group's normative ways of processing the world and from one another in their own idiosyncratic ways. Overall, our results suggest that a similar understanding of the world, as reflected in similar brain responses across people, may help humans achieve and maintain social connections.

**Methods**

**Characterization of the social networks in the two residential communities.** A total of 119 participants completed our social-network survey, with $n_{\text{residential community 1}} = 70$ and $n_{\text{residential community 2}} = 49$ people in the two residential communities. All participants were living in one of these two communities of a first-year dormitory in a large state university (University of California, Los Angeles) in the United States. All participants provided informed consent for the social-network survey in accordance with the Institutional Review Board of the University of California, Los Angeles. It was administered during December and January of the students' first year in the university, which began in the last week of September. Therefore, the subjects had been living together in their communities for 3–4 months prior to completing the social-network survey. In the survey, participants were first asked to indicate their full names and any nicknames by which they were known. This allowed us to match individuals' names with the number of friendship nominations that they received from other residents of their community. Participants were then asked to type the names of other people in their residential community with whom they interacted regularly. Participants answered the following prompt: "Consider the people you like to spend your free time with. Since you arrived at [institution name], who are the people you've socialized with most often? (Examples: eat meals with, hang out with, study with, spend time with)." The participants in the study could name as many people as they wished who fit that description without any restrictions, and no time limit was imposed on the survey. We adapted this question from prior research that investigated social networks of university students[13,24,50].

We used the IGRAPH package[51] in R[52] to analyze the social-network data. We constructed two networks (i.e., one for each residential community) and encoded the participants' answers with unweighted and directed edges. We then calculated the in-degree centrality of each individual. This quantity gives the number of the individual's community members (who participated in the social-network survey) who named them as someone with whom they interacted regularly. The distributions of the in-degree centralities were similar across the subsets of the fMRI sample from each community (see Supplementary Fig. 1).

**Combining data across residential communities.** As we noted in the prior subsection, each participant in our study was living in one of two residential communities and we operationalized popularity by calculating in-degree centrality as the number of nominations that each participant received from peers who were living in the same residential community. Each residential community was relatively bounded, and residents were encouraged (e.g., via intra-community social activities) to form social connections within their community. To maximize statistical power, we compared the neural responses across all possible pairs of participants (i.e., dyads) in both residential communities and then related the ISCs to in-degree centrality values across all possible pairs of individuals, including ones who were living in different residential communities. It is possible that this approach may have diminished our capacity to detect relationships between neural similarity and popularity, depending on how much the link between popularity within communities and neural similarity is based on community-specific norms. However, both communities consisted of first-year students who were attending the same university, so we reasoned that norms were likely to be similar across the two communities and that it would thus be reasonable for our neural analysis to include ISCs between individuals from different residential communities. We later complemented these main analyses with analyses that were based on only intra-community neural similarities. The results of these subsequent analyses (see Supplementary Figs. 3–4 and 8–9) yield similar results as our main analyses.

**fMRI study subjects**. A total of 70 participants from the aforementioned two residential communities participated in the neuroimaging portion of our study. We excluded two individuals due to excessive movement in more than half of the scan and excluded one individual who fell asleep during half of the scan. We also excluded one individual who did not complete either the scan or the social-network survey. Three additional fMRI participants did not complete the social-network survey. (See Supplementary Table 9 for a table of excluded participants.) This resulted in a total of 63 participants (40 female) between the ages of 18 and 21 (with a mean age of $M = 18.19$ and a standard deviation of $SD = 0.59$) that we included for all analyses. The distributions of in-degree centralities were similar across the fMRI participants and the full set of participants (see Supplementary Fig. 1). We included partial data from two fMRI participants. One participant had excessive head movement in one of the four runs, and one participant reported falling asleep in one of the four runs. In analyses that involved brain data, we excluded the associated runs for these participants and only included the remaining three runs for these subjects. All participants provided informed consent in accordance with the procedures of the Institutional Review Board of the University of California, Los Angeles.

**fMRI Procedure.** Participants attended an in-person study session that included self-report surveys and a 90-minute neuroimaging session in which we measured their brain activity using blood-oxygen-level-dependent (BOLD) fMRI. The fMRI data collection occurred between September and early November during the participants' first year at the university, and it was thus completed before the start of data collection for the social-network part of our study. Prior to entering the scanner, participants completed self-report surveys in which they provided demographic information, including their age, gender, and ethnicity. During the fMRI portion of the study, the participants watched 14 video clips with sound. The stimuli consisted of 14 different videos that varied in both duration (from 91 to 734 seconds) and content. (See Supplementary Table 1 for descriptions of the content.) Prior to scanning, we informed the participants that they would be watching video clips of heterogeneous content and that their experience would be like watching television while someone else "channel surfed"[vi]. The video clips were presented across four runs (as described in Supplementary Table 1) without breaks between clips within each run.

Some of the video clips have been used previously (10 of the videos were used in prior studies, and 4 of them are new), and we used similar criteria to those in prior work to select new stimuli[24,39]. First, we selected stimuli that were not likely to have been seen previously by the participants in an effort to avoid inducing inter-subject differences that arose from familiarity with the content. Second, we selected stimuli that were likely to be engaging to minimize the likelihood that participants would mind-wander during viewing, as this could potentially introduce undesirable noise into our data. Third, we selected stimuli that were likely to elicit substantial variability in the interpretations and meaning that different individuals can draw from the content. The participants were asked to watch the videos naturally (i.e., as they would watch them in a normal situation in life). All participants saw the videos in the same order to avoid any potential variability in neural responses from differences in the way that the stimuli were presented (rather than from endogenous individual-level differences). One can think of our consistently-ordered series of stimuli as a single continuous stream of content (analogous to different scenes in a movie). It is possible that different relative orderings of the stimuli could generate different results, similar to how reordering scenes in a movie might generate different results (e.g., due to differences in how tone, narratives, and themes evolve over the span of the

---

[vi] The term "channel-surfing" is an idiom that refers to scanning through different television channels to find something to watch.

movie). In our study, we presented the videos in the same order to all participants in order to keep the context surrounding each video consistent across participants because our main priority was to maximize sensitivity to individual-level differences. The video "task" was divided into four runs, and the task lasted approximately 60 minutes in total. Structural images of the brain were also collected. (We describe the image collection in more detail in the "fMRI data acquisition" subsection.) After the fMRI scan, the participants provided ratings (in the form of integers between 1 and 5) both on how much they enjoyed each video ("How much did you enjoy this video?"; response options ranged from 1 to 5, with the anchors "1 = not at all" and "5 = very much") and on how interesting they found each video ("How interesting did you find this video?"; response options ranged from 1 to 5, with the anchors "1 = very boring" and "5 = very interesting"). We obtained these preference ratings after the fMRI scan in an effort to minimize potential biases or disruptions in processing that could occur if participants were asked to reflect on content immediately after each stimulus was presented during scanning.

**fMRI data acquisition.** The participants were scanned using a 3T Siemens Prisma scanner with a 32-channel coil. Functional images were recorded using an echo-planar sequence (with echo time = 37 ms, repetition time = 800 ms, voxel size = 2.0 mm × 2.0 mm × 2.0 mm, matrix size = 104 × 104 mm, field of view = 208 mm, slice thickness = 2.0 mm, multi-band acceleration factor = 8, and 72 interleaved slices with no gap). A black screen was included at the beginning (with duration = 8 seconds) and the end (duration = 20 seconds) of each run to allow the BOLD signal to stabilize. We also acquired high-resolution T1-weighted (T1w) images (with echo time = 2.48 ms, repetition time = 1,900 ms, voxel size = 1.0 mm × 1.0 mm × 1.00 mm, matrix size = 256 × 256 mm, field of view = 256 mm, slice thickness = 1.0 mm, and 208 interleaved slices with 0.5 mm gap) for coregistration and normalization. We attached adhesive tape to the head coil in the MRI scanner and applied it across the participants' foreheads; it is known that this significantly reduces head motion[53].

**fMRI data analysis.** We used fMRIPrep version 1.4.0 for the data processing of our fMRI data[54]. We have taken the descriptions of anatomical and functional data preprocessing that begins in the next paragraph from the recommended boilerplate text that is generated by fMRIPrep and released under a CC0 license, with the intention that researchers reuse the text to facilitate clear and consistent descriptions of preprocessing steps, thereby enhancing the reproducibility of studies.

For each subject, the T1-weighted (T1w) image was corrected for intensity non-uniformity (INU) with N4BiasFieldCorrection, distributed with ANTs 2.1.0[55], and used as T1w-reference throughout the workflow. Brain tissue segmentation of cerebrospinal fluid (CSF), white matter (WM), and gray matter (GM) was performed on the brain-extracted T1w using FSL fast[56]. Volume-based spatial normalization to the ICBM 152 Nonlinear Asymmetrical template version 2009c (MNI152NLin2009cAsym) was performed through nonlinear registration with antsRegistration (ANTs 2.1.0[55]).

For each of the four BOLD runs per participant, the following preprocessing was performed. First, a reference volume and its skull-stripped version were generated using a custom methodology of fMRIPrep. The BOLD reference was then coregistered to the T1w reference using FSL flirt[56] with the boundary-based registration cost function. The coregistration was configured with nine degrees of freedom to account for distortions remaining in the BOLD reference. Head-motion parameters with respect to the BOLD reference (transformation matrices, and six corresponding rotation and translation parameters) were estimated before any spatiotemporal filtering using FSL mcflirt[56]. Automatic removal of motion artifacts using independent component analysis (ICA–AROMA) was performed on the preprocessed BOLD on MNI-space time series after removal of non-steady-state volumes and spatial smoothing with an isotropic, Gaussian kernel of 6mm FWHM (full-width half-maximum). The BOLD time series were then resampled to the MNI152Nlin2009cAsym standard space.

The following 10 confounding variables generated by fMRIPrep were included as nuisance regressors: global signals extracted from within the cerebrospinal fluid, white matter, and whole-brain masks, framewise displacement, three translational motion parameters, and three rotational motion parameters.

**Cortical parcellation into brain regions.** We extracted neural responses across the whole brain using the 200-parcel cortical parcellation scheme of Schaefer et al.[36] and 14 subcortical regions using the Harvard–Oxford subcortical atlas[37]. Together, this resulted in 214 regions that span the whole brain.

**Inter-subject correlations.** We extracted and concatenated preprocessed time-series data across all four runs for each subject, except for the two subjects for whom we used only partial data. For these two subjects, we concatenated their three usable runs into a single time series and then calculated ISCs for these subjects by comparing their data to the corresponding three runs in the other subjects. We extracted the mean time series in each of the 214 brain regions for each subject at each time point [i.e., at each repetition time (TR)]. Our analyses included 63 subjects after the various exclusions, so there were 1,952 unique dyads. For each unique dyad, we calculated the Pearson correlation between the mean time series of the neural response in each of the 214 brain regions. We then Fisher z-transformed the Pearson correlations and normalized the subsequent values (i.e., using z-scores) within each brain region.

**Subject-level analysis.** As we explained in the "Results" section, we were interested in whether an individual's in-degree centrality is associated with their mean neural similarity with their peers. To test this relationship, we transformed the dyad-level neural similarity measures into individual-level measures to obtain a single number that encoded an individual's mean neural similarity with other individuals for each brain region. For each individual, we calculated the mean Fisher *z*-transformed ISC value for them with each other individual in each brain region. We then fit a separate GLM for each brain region to test the association between individual differences in in-degree centrality and the mean neural similarity in the respective brain region. We FDR-corrected all results because of the multiple comparisons.

**Dyad-level ISC analysis**. For our dyad-level analysis, we took the following steps to test for associations between in-degree centrality and neural similarity in each of the 214 brain regions. First, we transformed the subject-level in-degree centrality measure into a dyad-level measure by creating a binarized variable that indicated whether the two members of the dyad had high, low, or mixed in-degree centralities (i.e., {high, high}, {low, low}, or {low, high}). See the "Results" section for details. Of the 1,952 unique dyads, 253 of them were {high, high}, 779 of them were {low, low}, and 920 of them were {low, high}. To relate this dyad-level in-degree centrality measure and neural similarity, we used the method in Chen et al.[40] and fit linear mixed-effects models with crossed random effects using LME4[57] and LMERTEST[57] in R. This approach allowed us to account for non-independence in the data from the repeated observations for each subject (i.e., because each subject is part of multiple dyads). Following the method that was suggested by Chen et al.[40], we "doubled" the data (adding redundancy) to allow fully-crossed random effects. In other words, we accounted for the symmetric nature of the ISC matrix and the fact that one participant contributes twice in a dyad (i.e., $(i, j) = (j, i)$ for participants *i* and $j$). See Chen et al.[40] for more details. Following Chen et al.[40], we manually corrected the degrees of freedom to $N - k$, where $N$ is the number of unique observations (in our case, $N = 1,952$ because there are 1,952 unique dyads) and $k$ is the number of fixed effects in the model, before performing statistical inference. All findings that we report in the present paper use the corrected number of degrees of freedom. For each brain region, we first fit a mixed-effects model to infer neural similarity in that brain region from the binarized dyad-level in-degree variable, with random intercepts for each member of the dyad (i.e., "subject 1" and "subject 2"). We then conducted planned-contrast analyses using EMMEANS[58] in R to compare which brain regions had larger ISCs for the different values of the dyadic in-degree centrality variable: $ISC_{\{high, high\}} > ISC_{\{low, low\}}$, $ISC_{\{high, high\}} > ISC_{\{low, high\}}$, and $ISC_{\{low, high\}} > ISC_{\{low, low\}}$. We transformed all variables into z-scores prior to our subsequent computations to obtain standardized coefficients ($B$) as outputs. We FDR-corrected all $p$-values at $p < 0.001$ because of multiple comparisons.

**Dyad-level behavioral analysis**. We took an analogous approach as in our dyad-level ISC analysis to test the relationships between dyadic in-degree centrality and preference similarity. (See "Preference similarity" in the "Results" section.) To do this, we followed the same procedure as the one that we described above in "Dyad-level ISC analysis" and fit two mixed-effects models that take into account the dependency structure of the data. We constructed one such model for each type of rating (i.e., enjoyment and interest) to infer a similarity from the dyad-level in-degree variable, with random intercepts for each member of the dyad. We then conducted planned-contrast analyses using EMMEANS[58] in R to examine whether there was an association between preference similarity and different levels of the dyadic in-degree centrality variable: $s_{\{high, high\}} > s_{\{low, low\}}$, $s_{\{high, high\}} > s_{\{low, high\}}$, and $s_{\{low, high\}} > s_{\{low, low\}}$, where $s$ is the dyad-level preference similarity that we defined in our discussion of "Preference similarity". We transformed all variables into *z*-scores prior to our subsequent calculations to obtain standardized coefficients (*B*) as outputs. We used an FDR-corrected significance threshold of $p < 0.001$ because of the multiple comparisons from the planned contrasts.

Data availability

The data that support the findings of this study are available from the corresponding author upon reasonable request.

Code availability

The code used for the analyses is available from the corresponding author upon reasonable request.

*for Statistical Computing* (2013).

53. Krause, F. *et al.* Active head motion reduction in magnetic resonance imaging using tactile feedback. *Hum. Brain Mapp.* **40**, 4026–4037 (2019).

54. Esteban, O. *et al.* FMRIPrep: A robust preprocessing pipeline for functional MRI. *Nat. Methods* **16**, 111–116 (2019).

55. Avants, B. B. *et al.* A reproducible evaluation of ANTs similarity metric performance in brain image registration. *NeuroImage* **54**, 2033–2044 (2011).

56. Smith, S. M. *et al.* Advances in functional and structural MR image analysis and implementation as FSL. *NeuroImage* **23**, S208–S219 (2004).

57. Kuznetsova, A., Brockhoff, P. B. & Christensen, R. H. B. lmerTest package: Tests in linear mixed effects models. *J. Stat. Softw.* **82**, (2017).

58. Russell, A. *et al.* EMMEANS: Estimated Marginal Means, aka Least-Squares Means. *CRAN* (2021) doi:10.1080/00031305.1980.10483031.45

Acknowledgements

We thank Elena Sternlicht, Kelly Xue, and the UCLA Center for Cognitive Neuroscience (particularly Jared Gilbert) for providing support with data collection. We thank Oshton Tsen for providing support with the figures. This work was supported by the National Science Foundation SBE Postdoctoral Research Fellowship [Grant No. 1911783] and the National Science Foundation [Grant No. SBE-1835239].

Author Contributions

E.C.B., R.H., M.A.P., and C.P. designed the study and experiments. E.C.B., R.H., and K.L. collected the data. E.C.B. analyzed the data with support from R.H., E.S.F., and C.P. E.C.B., M.A.P., and C.P. wrote the manuscript with feedback from all authors.

Competing Interests

The authors declare no competing interests.

Supplementary Information for **"Popular individuals process the world in particularly normative ways"**

Elisa C. Baek[1*], Ryan Hyon[1], Karina López[1], Emily S. Finn[2], Mason A. Porter[3,4], and Carolyn Parkinson*[1,5]

[1]Department of Psychology, University of California, Los Angeles, [2]Department of Psychological and Brain Sciences, Dartmouth College, [3] Department of Mathematics, University of California, Los Angeles, [4]Santa Fe Institute, [5] Brain Research Institute, University of California, Los Angeles

* Corresponding authors

## Supplementary Fig. 1 for "Social-network characterization" in the "Results" section: Distribution of in-degree centrality

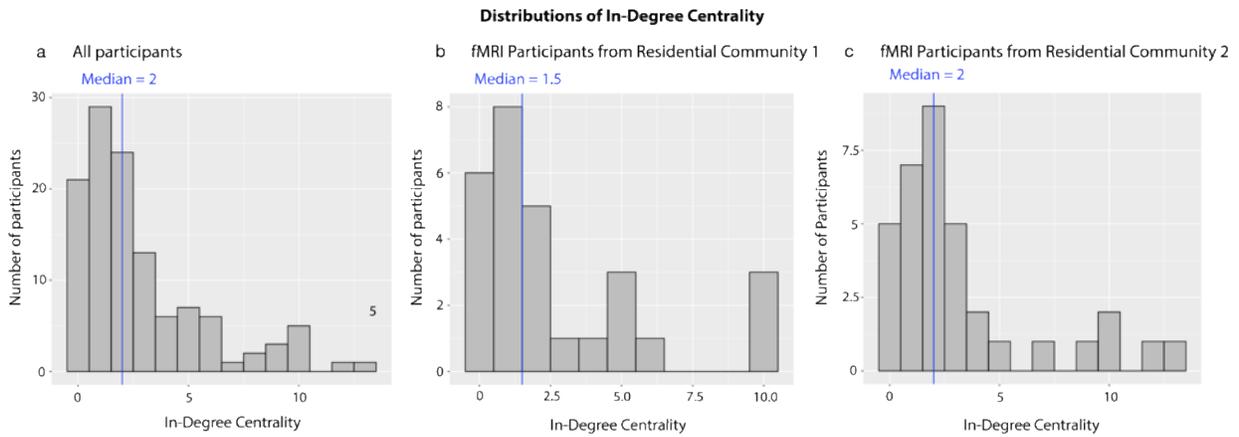

**Supplementary Fig. 1.** Distributions of in-degree centrality. The distributions of in-degree centrality are relatively similar across the two residential communities. (a) The distribution of in-degree centrality for the set of all participants who participated in our social-network survey. (b) The distribution of in-degree centrality in the subset of the fMRI participants who were living in residential community 1. (c) The distribution of in-degree centrality in the subset of the fMRI participants who were living in residential community 2.

# Supplementary Table 1 for "Neural similarity" in the "Results" section: Stimuli descriptions

Supplementary Table 1. Descriptions of stimuli

| Video # | Run # | Video | Content |
|---|---|---|---|
| 1 | 1 | An Astronaut's View of Earth | An astronaut discusses viewing Earth from space and, in particular, witnessing the effects of climate change from space. He then urges viewers to mobilize to address this issue. |
| 2 | 1 | All I Want | A sentimental music video depicting a social outcast with a facial deformity who is seeking companionship. |
| 3 | 1 | Scientific demonstration | An astronaut at the International Space Station demonstrates and explains what happens when one wrings out a waterlogged washcloth in space. |
| 4 | 1 | Food Inc. | An excerpt from a documentary discussing how the fast-food industry influences food production and farming practices in the United States. |
| 5 | 2 | We Can Be Heroes | An excerpt from a mockumentary-style series in which a man discusses why he nominated himself for the title of Australian of the Year. |
| 6 | 2 | Ban College Football | Journalists and athletes debate whether football should be banned as a college sport. |
| 7 | 2 | Soccer match | Highlights from a soccer match. |
| 8 | 2 | Ew! | A comedy skit in which grown men play teenage girls disgusted by the things around them. |
| 9 | 2 | Life's Too Short | An example of a 'cringe comedy' in which a dramatic actor is depicted unsuccessfully trying his hand at improvisational comedy. |
| 10 | 2 | America's Funniest Home Videos | A series of homemade video clips that depict examples of unintentional physical comedy arising from accidents. |
| 11 | 3 | Zima Blue | A philosophical, animated short set in a futuristic world. |
| 12 | 3 | Nathan For You | An episode from a 'docu-reality' comedy in which the host convinces people, who are not always in on the joke, to engage in a variety of strange behaviors. |
| 13 | 4 | College Party | An excerpt from a film that depicts a party scene in which a bashful college student is pressured to drink alcohol. |
| 14 | 4 | Eighth Grade | Two excerpts from a film. They depict (1) a young teenager who video blogs about her mental-health issues and (2) an awkward scene between two teenagers on a dinner date. |

Note: Videos 1–10 are a subset of the videos that were used in a prior study[1]. The descriptions of them in the present paper are the same as those in the prior study.

# Supplementary Table 2 and Supplementary Table 3 for "Subject-level ISC analysis" in the "Results" section: Subject-level subcortical results

Supplementary Table 2. Subject-level results that relate ISCs with the binarized in-degree centrality (high versus low): Subcortical results

| Subcortical region | B | SE | p (corrected) | p (uncorrected) |
|---|---|---|---|---|
| Accumbens (L) | 0.363 | 0.260 | 0.325 | 0.167 |
| Amygdala (L) | 0.578 | 0.253 | 0.164 | 0.026 |
| Caudate (L) | 0.269 | 0.262 | 0.466 | 0.307 |
| Hippocampus (L) | 0.550 | 0.254 | 0.180 | 0.031 |
| Pallidum (L) | 0.611 | 0.252 | 0.144 | 0.018 |
| Putamen (L) | 0.300 | 0.261 | 0.415 | 0.256 |
| Thalamus (L) | 0.432 | 0.258 | 0.247 | 0.099 |
| Accumbens (R) | 0.080 | 0.264 | 0.834 | 0.764 |
| Amygdala (R) | 0.493 | 0.256 | 0.210 | 0.059 |
| Caudate (R) | 0.171 | 0.263 | 0.640 | 0.519 |
| Hippocampus (R) | 0.587 | 0.253 | 0.158 | 0.024 |
| Pallidum (R) | 0.068 | 0.264 | 0.845 | 0.798 |
| Putamen (R) | 0.302 | 0.261 | 0.414 | 0.252 |
| Thalamus (R) | 0.482 | 0.256 | 0.220 | 0.065 |

We have FDR-corrected all $p$-values because of multiple comparisons; we also report the corresponding uncorrected $p$-values. The quantity $B$ is the standardized regression coefficient, and the quantity SE is the standard error.

Supplementary Table 3. Subject-level results that relate ISCs with the original (i.e., non-binarized) in-degree centrality: Subcortical results

| Subcortical region | $\rho$ | p (corrected) | p (uncorrected) |
|---|---|---|---|
| Accumbens (L) | 0.248 | 0.217 | 0.050 |
| Amygdala (L) | 0.350 | 0.104 | 0.005 |
| Caudate (L) | 0.167 | 0.355 | 0.191 |
| Hippocampus (L) | 0.302 | 0.175 | 0.016 |
| Pallidum (L) | 0.256 | 0.217 | 0.043 |
| Putamen (L) | 0.123 | 0.491 | 0.337 |
| Thalamus (L) | 0.212 | 0.262 | 0.096 |
| Accumbens (R) | 0.054 | 0.787 | 0.673 |
| Amygdala (R) | 0.217 | 0.262 | 0.088 |
| Caudate (R) | 0.063 | 0.742 | 0.624 |
| Hippocampus (R) | 0.279 | 0.200 | 0.027 |
| Pallidum (R) | –0.003 | 0.989 | 0.981 |
| Putamen (R) | 0.154 | 0.392 | 0.227 |
| Thalamus (R) | 0.251 | 0.217 | 0.047 |

We have FDR-corrected all $p$-values because of multiple comparisons; we also report the corresponding uncorrected $p$-values.

**Supplementary Methods 1 for "Subject-level ISC analysis" in the "Results" section**

**Binarized subject-level results that control for similarities in self-reported demographic traits.** We fit analogous models to those that we described in the "Subject-level ISC analysis" in the "Results" section of the main manuscript to test the relationship between in-degree centrality and ISC while controlling for all available self-reported demographic variables: similarities in age, gender, ethnicity, and home country (which we define as the country where an individual was living prior to enrolling at the university). To control for similarities in demographic variables, for each unique dyad (i.e., for each pair of individuals) who participated in the fMRI session, we computed an absolute difference of the age between each individual in the dyad (i.e., age_difference = |$age_1$ – $age_2$|). We then transformed this difference score into a similarity score such that larger numbers indicated greater similarity (specifically, age_similarity = 1 – (age_difference/max(age_difference)). To control for similarities in gender, we created an indicator variable in which 0 signifies different genders and 1 signifies the same gender. To control for similarities in ethnicity and race, we created an indicator variable for each ethnicity category (Asian, Black/African, Hispanic/Latinx, Native American, Pacific Islander, and Caucasian/White) in which 0 signifies a different self-reported ethnicity and 1 signifies the same self-reported ethnicity. Participants were able to self-report as many ethnicities as they desired. For each unique dyad, we used an overall indicator variable for ethnicity in which 0 signifies no shared ethnicity and 1 signifies a shared ethnicity. If two members of a dyad self-reported even one same ethnicity, we coded them as having a shared ethnicity. To control for similarities in home country, we used an additional indicator variable in which 0 signifies different home countries and 1 signifies the same home country. For each subject, we then calculated the mean similarity between them and each other participant for each of the demographic variables and used these variables as covariates in our mixed-effects models. These models gave a similar pattern of results as those that we reported in the main manuscript (see Supplementary Fig. 2).

**Supplementary Figure 2 for "Subject-level ISC analysis" in the "Results" section**

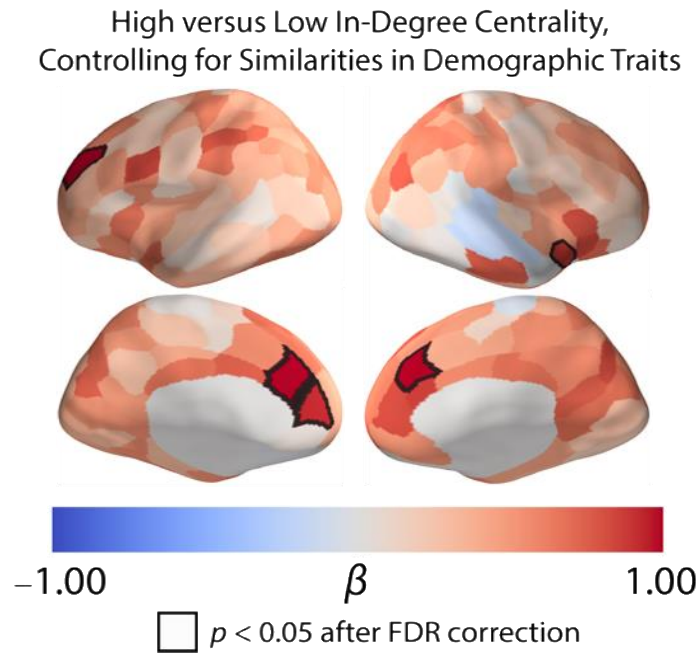

**Supplementary Fig. 2.** Subject-level results that control for similarities in demographic traits (specifically, for mean similarities with other participants in age, gender, ethnicity and race, and home country). The association between in-degree centrality and ISCs in the bilateral DMPFC that we reported in the main manuscript remains significant after controlling for similarities in demographic traits. The quantity $B$ is the standardized regression coefficient. Regions where we observed significant associations between in-degree centrality and ISC are outlined in black. We used an FDR-corrected significance threshold of $p < 0.05$, which corresponds to an uncorrected $p$-value threshold of $p < 0.001$.

**Supplementary Methods 2 for "Subject-level ISC analysis" in the "Results" section**

**Binarized subject-level results: Intra-community ISC only.** Because our participants came from two different residential communities, we also conducted analyses to test if we could observe similar patterns in associations between ISCs and popularity while only using intra-community ISCs. To do this, we fit analogous models as to those that we described in the "Subject-level ISC analysis" in the "Results" section of the main manuscript, except that we calculated each subject's mean neural similarity in each brain region using only other members of their own residential community. That is, we removed dyads in which both members did not belong to the same community. The results of these calculations (see Supplementary Fig. 3) are similar to those that we reported in the main manuscript.

**Supplementary Fig. 3 for "Subject-level ISC analysis" in the "Results" section**

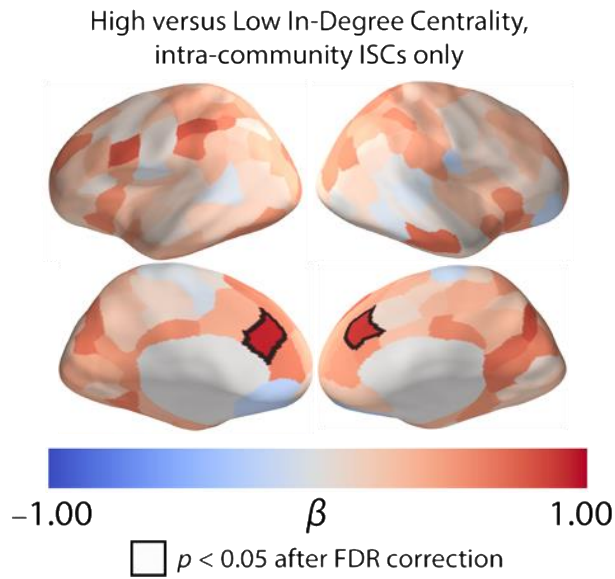

**Supplementary Fig. 3.** Subject-level results for intra-community ISCs only. The results of our calculations using only intra-community ISCs identified a similar pattern of results (specifically, large ISCs in the bilateral DMPFC) as in our main analysis (see Fig. 2) as significantly associated with in-degree centrality. The quantity *B* is the standardized regression coefficient. Regions where we observed significant associations between in-degree

centrality and ISC are outlined in black. We used an FDR-corrected significance threshold of $p < 0.05$, which corresponds to an uncorrected $p$-value threshold of $p < 0.001$.

**Supplementary Methods 3 for "Subject-level ISC analysis" in the "Results" section**

**Binarized subject-level results: Intra-community ISCs only that control for self-reported demographic traits and social distances.** We also conducted analyses to test if the associations between ISCs and in-degree centrality remain significant after we control both for demographic variables (specifically, for mean similarities with community members in age, gender, ethnicity and race, and home country[vii]) and for the social distances between individuals, given that social distance has been associated previously with neural similarities [1,2]. To do this, we fit GLMs to infer ISCs in each brain region from in-degree centrality while including all of the control variables as covariates in the model. For these computations, we used ISCs in intra-community dyads only, so we calculated each individual's mean ISC based on their ISCs only with other members of their own residential community. For each individual, we took the following steps to calculate their mean social distance from each other participant from their community. For each intra-community dyad, we defined "social distance" to be the smallest number of intermediate social ties (i.e., the geodesic distance) that is necessary to connect the two individuals in the network via a path. We calculated social distances using an unweighted and undirected network that includes a tie between two individuals even if only one of them nominated the other as a friend. We set the social distance between any pair of individuals who were on different components of the network as equal to 1 more than the maximum measured social distance in the network. In both of the residential communities, the maximum measured social distance was 6, so we coded such individuals who were on different components of the

---

[vii]See "Binarized subject-level results that control for similarities in self-reported demographic traits" above for details about how we calculated these similarities.

network to have a social distance of 7 between them. In our subject-level analysis, for each individual, we took the mean social distance between them and each other member of their residential community and used this variable as a covariate in our GLM for inferring ISCs. The results of these calculations (see Supplementary Fig. 4) are similar to those that we reported in the main manuscript.

**Supplementary Fig. 4 for "Subject-level ISC analysis" in the "Results" section**

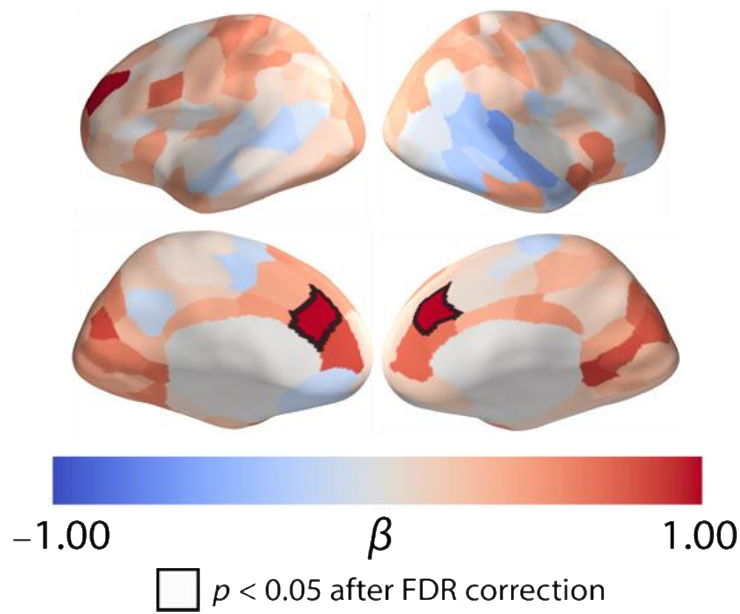

**Supplementary Fig. 4.** Subject-level results for intra-community ISCs only when controlling both for similarities in demographic traits and for social distances. We again identified a similar pattern of results (specifically, large ISCs in the bilateral DMPFC) as in our main analysis (see Fig. 2) as significantly associated with in-degree centrality. The quantity $B$ is the standardized regression coefficient. Regions where we observed significant associations between in-degree centrality and ISC are outlined in black. We used an FDR-corrected significance threshold of $p < 0.05$. which corresponds to an uncorrected $p$-value threshold of $p < 0.001$.

**Supplementary Methods 4 for "Subject-level ISC analysis" in the "Results" section**

**Binarized subject-level results with approximately equal-sized groups.** Because our median-split approach yields unequally sized in-degree centrality groups ($n_{high} = 23$; $n_{low} = 40$), we also tested associations between ISCs and the binarized in-degree centrality variable with approximately equal-sized groups. To do so, we took a subset of the data by removing all participants with the median in-degree centrality value (specifically, with an in-degree centrality equal to 2). We then classified participants as part of the high-centrality group if they had an in-degree that was larger than the median (specifically, if it was more than 2; there were $n_{high} = 23$ such people) and into the low-centrality group if they had an in-degree that was less than the median (specifically, if it was less than 2; there were $n_{low} = 26$ such people). Similar to our main subject-level results, relationships between ISCs in the right and left DMPFC and in-degree centrality were significant even when using this subset of the data. We also found that the ISCs in other regions (including the precuneus and the inferior parietal lobule) of the default-mode network were also associated with in-degree centrality (see Supplementary Fig. 5).

**Supplementary Fig. 5 for "Subject-level ISC analysis" in the "Results" section**

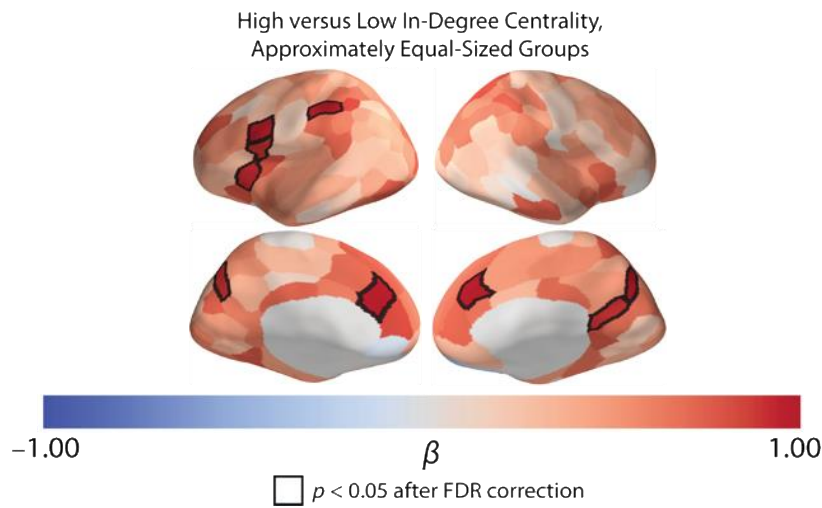

**Supplementary Fig. 5.** Subject-level results with approximately equal-sized in-degree centrality groups. The associations between ISCs in the right and left DMPFC and in-degree centrality remain significant when we used a subset of the data with approximately equal-sized in-degree centrality groups. Additionally, we found that the ISCs in other regions (including the precuneus and the inferior parietal lobule) of the default-mode network were also associated with in-degree centrality. The quantity *B* is the standardized regression coefficient. Regions where we observed significant associations between in-degree centrality and ISC are outlined in black. We used an FDR-corrected significance threshold of $p < 0.05$, which corresponds to an uncorrected *p*-value threshold of $p < 0.002$.

**Supplementary Methods 5 for "Subject-level ISC analysis" in the "Results" section**

**Binarized subject-level results that control for preference similarities.** The relationships between in-degree centrality and the ISCs in the right and left DMPFC remain significant after controlling for similarities in enjoyment and interest ratings. This suggests that neural similarities in these regions capture similarities beyond those that we observed from self-reported preference ratings (see Supplementary Fig. 6).

**Supplementary Fig. 6 for "Subject-level ISC analysis" in the "Results" section**

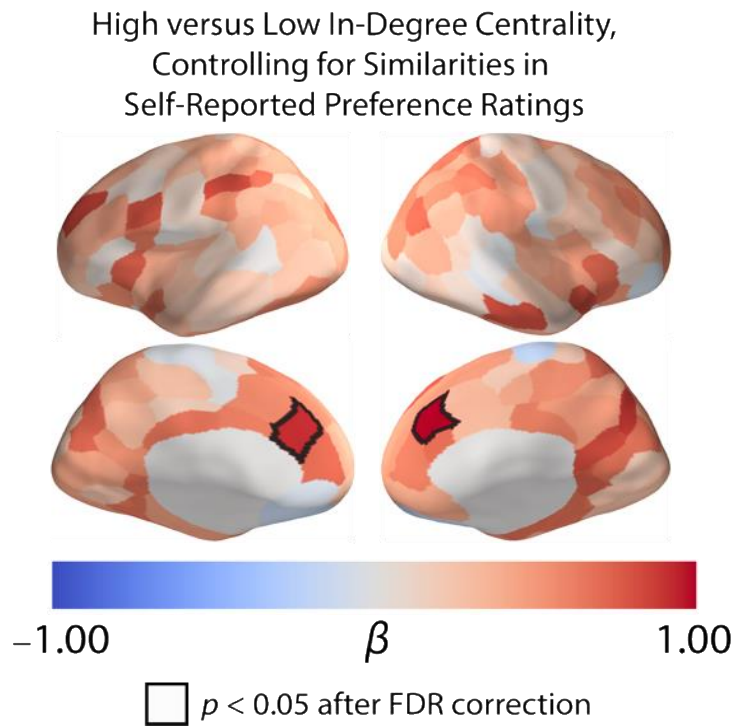

**Supplementary Fig. 6.** Subject-level results that control for similarities in self-reported preference ratings. The associations between in-degree centrality and ISCs in the right and left DMPFC and remain significant after controlling for similarities in self-reported preference ratings (i.e., ratings of enjoyment of and interest in stimuli). The quantity $B$ is the standardized regression coefficient. Regions where we observed significant associations between in-degree centrality and ISC are outlined in black. We used an FDR-corrected significance threshold of $p < 0.05$, which corresponds to an uncorrected $p$-value threshold of $p < 0.001$.

# Supplementary Tables 4–6 for "Dyad-level ISC analysis" in the "Results" section: Subcortical results

Supplementary Table 4. Dyad-level results that relate ISCs with binarized in-degree centrality
Contrast: $ISC_{\{high, high\}} > ISC_{\{low, low\}}$

| Subcortical region | B | SE | p (corrected) | p (uncorrected) |
|---|---|---|---|---|
| Accumbens (L) | 0.194 | 0.120 | 0.054 | 0.022 |
| Amygdala (L) | 0.505 | 0.218 | 0.007* | 0.001 |
| Caudate (L) | 0.193 | 0.174 | 0.190 | 0.117 |
| Hippocampus (L) | 0.448 | 0.208 | 0.012* | 0.002 |
| Pallidum (L) | 0.263 | 0.109 | 0.005** | < 0.001 |
| Putamen (L) | 0.185 | 0.160 | 0.168 | 0.107 |
| Thalamus (L) | 0.218 | 0.146 | 0.078 | 0.035 |
| Accumbens (R) | –0.026 | 0.114 | 0.795 | 0.749 |
| Amygdala (R) | 0.402 | 0.210 | 0.024* | 0.007 |
| Caudate (R) | 0.105 | 0.172 | 0.466 | 0.385 |
| Hippocampus (R) | 0.525 | 0.213 | 0.004** | < 0.001 |
| Pallidum (R) | 0.020 | 0.094 | 0.807 | 0.763 |
| Putamen (R) | 0.242 | 0.175 | 0.103 | 0.052 |
| Thalamus (R) | 0.294 | 0.171 | 0.041* | 0.015 |

\* $p < 0.05$, \*\* $p < 0.01$; we have FDR-corrected all $p$-values because of multiple comparisons.
We also report the corresponding uncorrected $p$-values.

Supplementary Table 5. Dyad-level results that relate ISCs with binarized in-degree centrality
Contrast: $ISC_{\{high, high\}} > ISC_{\{low, high\}}$

| Subcortical region | B | SE | p (corrected) | p (uncorrected) |
|---|---|---|---|---|
| Accumbens (L) | 0.151 | 0.011 | 0.034* | 0.011 |
| Amygdala (L) | 0.268 | 0.002 | 0.009** | 0.002 |
| Caudate (L) | 0.123 | 0.092 | 0.156 | 0.092 |
| Hippocampus (L) | 0.221 | 0.007 | 0.024* | 0.007 |
| Pallidum (L) | 0.139 | 0.015 | 0.041* | 0.015 |
| Putamen (L) | 0.099 | 0.151 | 0.232 | 0.151 |
| Thalamus (L) | 0.064 | 0.331 | 0.413 | 0.331 |
| Accumbens (R) | –0.120 | 0.039 | 0.085 | 0.039 |
| Amygdala (R) | 0.197 | 0.017 | 0.046* | 0.017 |
| Caudate (R) | 0.043 | 0.551 | 0.625 | 0.551 |
| Hippocampus (R) | 0.319 | 0.000 | 0.002** | < 0.001 |
| Pallidum (R) | 0.004 | 0.941 | 0.951 | 0.941 |
| Putamen (R) | 0.192 | 0.009 | 0.029* | 0.009 |
| Thalamus (R) | 0.101 | 0.162 | 0.241 | 0.162 |

\* $p < 0.05$, \*\* $p < 0.01$; we have FDR-corrected all $p$-values because of multiple comparisons.
We also report the corresponding uncorrected $p$-values.

Supplementary Table 6. Dyad-level results that relate ISCs with binarized in-degree centrality
Contrast: $ISC_{\{low, high\}} > ISC_{\{low, low\}}$

| Subcortical region | B | SE | p (corrected) | p (corrected) |
|---|---|---|---|---|
| Accumbens (L) | 0.043 | 0.365 | 0.448 | 0.364 |
| Amygdala (L) | 0.237 | 0.003 | 0.013* | 0.003 |
| Caudate (L) | 0.070 | 0.282 | 0.367 | 0.282 |
| Hippocampus (L) | 0.226 | 0.003 | 0.014* | 0.003 |
| Pallidum (L) | 0.124 | 0.006 | 0.022* | 0.006 |
| Putamen (L) | 0.086 | 0.153 | 0.233 | 0.153 |
| Thalamus (L) | 0.154 | 0.006 | 0.022* | 0.006 |
| Accumbens (R) | 0.094 | 0.041 | 0.088 | 0.041 |
| Amygdala (R) | 0.205 | 0.008 | 0.026* | 0.008 |
| Caudate (R) | 0.062 | 0.332 | 0.413 | 0.332 |
| Hippocampus (R) | 0.206 | 0.008 | 0.027* | 0.008 |
| Pallidum (R) | 0.016 | 0.692 | 0.746 | 0.692 |
| Putamen (R) | 0.050 | 0.445 | 0.518 | 0.445 |
| Thalamus (R) | 0.193 | 0.003 | 0.012* | 0.003 |

* $p < 0.05$, ** $p < 0.01$; we have FDR-corrected all $p$-values because of multiple comparisons. We also report the corresponding uncorrected $p$-values.

**Supplementary Methods 6 for "Dyad-level ISC analysis" in the "Results" section**

**Binarized dyad-level ISC results that control for similarities in demographic traits and friendships.** We also tested whether our findings remain significant after controlling for dyadic similarities in demographic traits (specifically, similarities in age, gender, ethnicity and race, and home country) and whether or not a given pair of participants were friends with each other. (See "Binarized subject-level results that control for similarities in self-reported demographic traits" above for descriptions of how we computed similarities in these variables.) If either individual in a dyad nominated the other as a friend, we coded the dyad as signifying an undirected friendship. We controlled for friendship, rather than social distance, because our primary analyses included ISCs between individuals from different residential communities. In "Binarized dyad-level results: Intra-community ISCs only" below, we discuss analyses that are based only on intra-community dyads that also control for social distances between individuals in a dyad. Our results indicate that our pattern of results that links ISC and dyad-level in-degree centrality remains significant after we control for demographic variables and friendship (see Fig. S6). This suggests that our findings that the greatest neural similarity occurs between highly-central individuals and that lower neural similarity occurs between less-central individuals arose from differences beyond what we captured using similarities in demographic traits and friendships between individuals in the same dyad.

**Supplementary Fig. 7 for "Dyad-level ISC analysis" in the "Results" section**

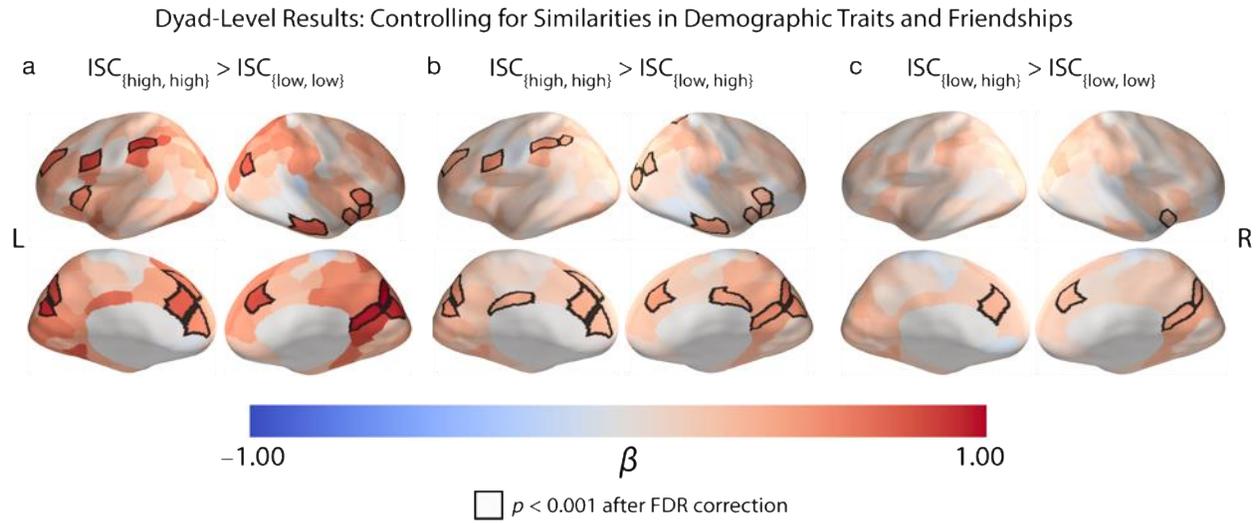

**Supplementary Fig. 7.** Dyad-level associations of neural similarity with in-degree centrality when controlling for similarities in demographic traits and friendships. We identified similar brain regions as significantly associated with in-degree centrality as in our results in the main manuscript (see Fig. 4). The quantity $B$ is the standardized regression coefficient. Regions where we observed significant associations between in-degree centrality and ISC are outlined in black. We used an FDR-corrected significance threshold of $p < 0.001$, which corresponds to an uncorrected $p$-value threshold of $p < 1.947 \times 10^{-6}$.

**Supplementary Methods 7 for "Dyad-level ISC analysis" in the "Results" section**

**Binarized dyad-level results: Intra-community ISCs only.** Because our participants came from two different residential communities, we also conducted analyses to test if our findings remain significant when we used only intra-community ISCs. To do this, we fit analogous models to those that we described in "Dyad-level ISC analysis" in the "Results" section of the main manuscript to test for associations between ISCs and dyad-level popularity, except that we calculated each subject's mean neural similarity in each brain region based only on other members of their own residential community. (That is, we removed dyads with individuals that did not belong to the same community.) We obtained similar results (see Supplementary Fig. 8) as the ones that we reported in the main manuscript.

**Supplementary Fig. 8 for "Dyad-level ISC analysis" in the "Results" section**

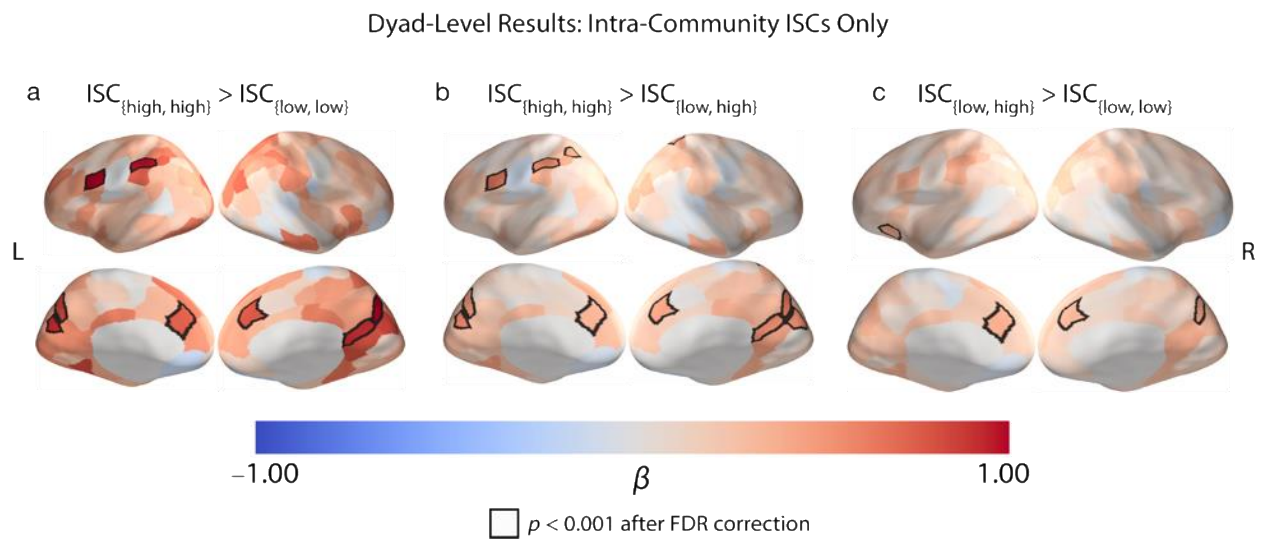

**Supplementary Fig. 8.** Dyad-level associations of neural similarity with in-degree centrality when we used intra-community ISCs only. We identified similar brain regions as significantly associated with in-degree centrality as in our results in the main manuscript (see Fig. 4). The quantity $B$ is the standardized regression coefficient. Regions where we observed significant associations between in-degree centrality and ISC are outlined in black. We used an FDR-corrected significance threshold of $p < 0.001$, which corresponds to an uncorrected $p$-value threshold of $p < 3.738 \times 10^{-5}$.

**Supplementary Methods 8 for "Dyad-level ISC analysis" in the "Results" section**

**Binarized dyad-level results: Intra-community ISCs only when controlling both for demographic variables and for social distances.** We also tested if our results remain significant after controlling both for demographic variables (specifically, for mean similarities with community members in age, gender, ethnicity and race, and home country[viii]) and for social distances between individuals. To do this, we fit mixed-effects models that infer ISCs in each brain region from the dyad-level in-degree centrality variable while including all of the control variables as covariates. For these analyses, we examined ISCs in intra-community dyads only, so we included only data from individuals from the same residential community. We defined social distance as we described in "Binarized dyad-level ISC results that control for similarities in demographic traits and friendships", and we included the social distance between the individuals of each dyad as a covariate in our mixed-effects models for inferring ISC in each brain region from the dyad-level in-degree centrality variable. We obtained results (see Supplementary Fig. 9) that were similar to those that we reported in the main manuscript. This suggests that our finding that neural similarity is linked with popularity is not merely a confound of the most-popular individuals being more likely to be friends with one another.

---

[viii]See "Binarized subject-level results that control for similarities in self-reported demographic traits" above for descriptions on how we computed similarities in these variables.

**Supplementary Fig. 9 for "Dyad-level ISC analysis" in the "Results" section**

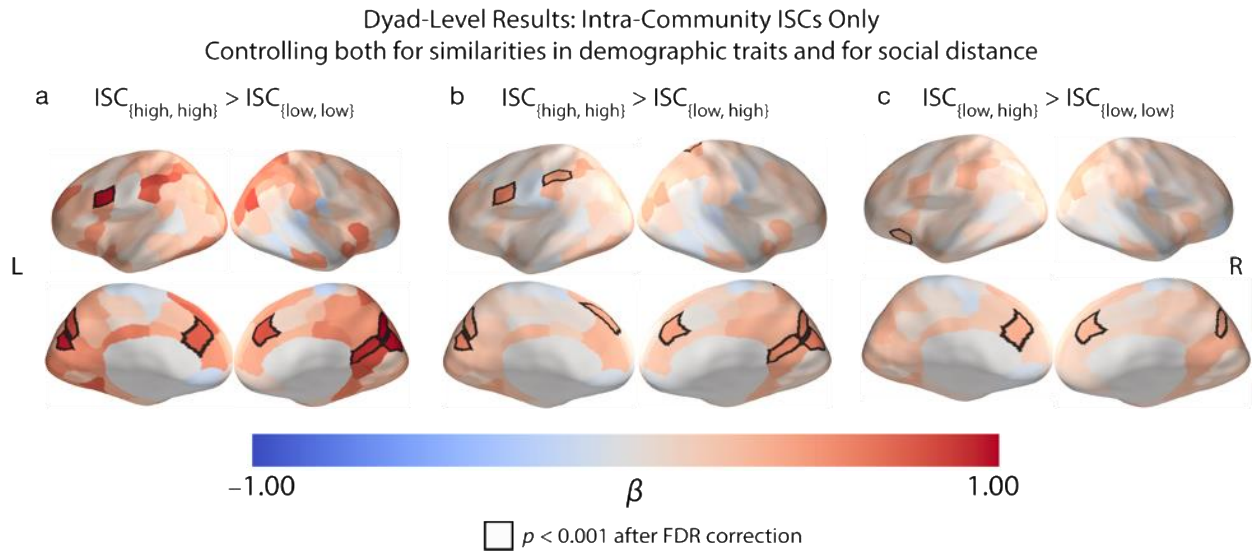

**Supplementary Fig. 9.** Dyad-level associations of neural similarity with in-degree centrality for intra-community ISCs only when controlling both for similarities in demographic traits and for social distances between individuals in a dyad. We identified similar brain regions as significantly associated with in-degree centrality as in our results in the main manuscript (see Fig. 4). The quantity $B$ is the standardized regression coefficient. Regions where we observed significant associations between in-degree centrality and ISC are outlined in black. We used an FDR-corrected significance threshold of $p < 0.001$, which corresponds to an uncorrected $p$-value threshold of $p < 3.583 \times 10^{-5}$.

**Supplementary Methods 9 for "Dyad-level ISC analysis" in the "Results" section**

**Binarized dyad-level results with approximately equal-sized groups.** Because our median-split approach yields unequally sized in-degree centrality groups, we also tested the associations between ISCs and the binarized in-degree centrality variable with approximately equal-sized groups of individuals (see Supplementary Methods 4). We took an analogous approach to one that we discussed in the main manuscript (see the Results and Methods sections for more details) to transform the subject-level in-degree centrality measure into a dyad-level measure. Of the 1,175 unique dyads, 324 of them were {high, high}, 253 of them were {low, low}, and 598 of them were {low, high}. We obtained results (see Supplementary Fig. 10) that were similar to those that we reported in the main manuscript.

**Supplementary Fig. 10 for "Dyad-level ISC analysis" in the "Results" section**

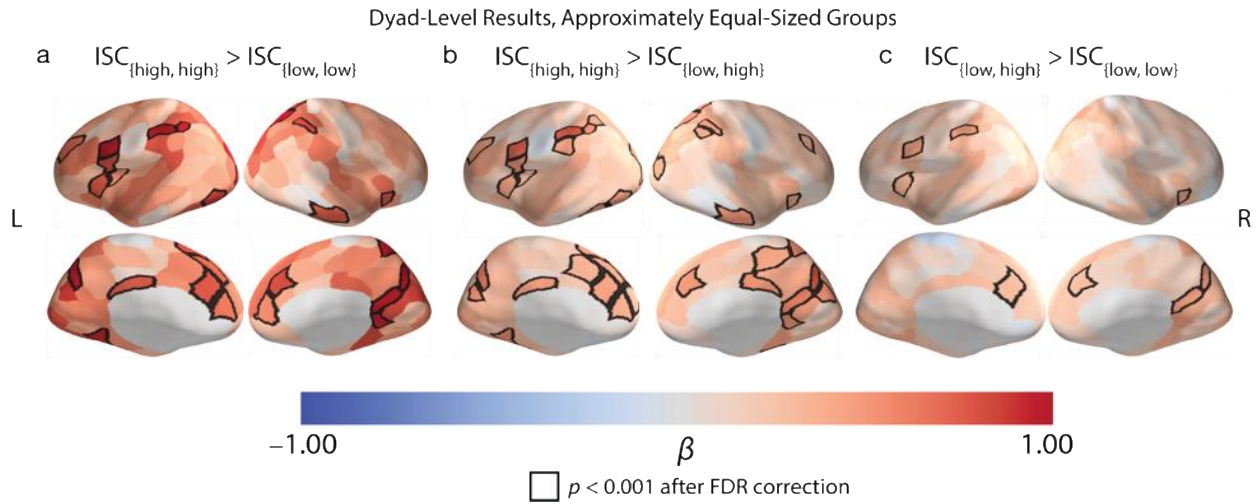

**Supplementary Fig. 10.** Dyad-level associations of neural similarity with in-degree centrality with approximately equal-sized in-degree centrality groups. We identified similar brain regions as significantly associated with in-degree centrality as in our results in the main manuscript (see Fig. 4) when we used a subset of the data with approximately equal-sized centrality groups. The quantity $B$ is the standardized regression coefficient. Regions where we observed significant associations between in-degree centrality and ISC are outlined in black. We used an FDR-corrected significance threshold of $p < 0.001$, which corresponds to an uncorrected $p$-value threshold of $p < 0.0001$.

**Supplementary Methods 9 for "Dyad-level preference analysis" in the "Results" section**

Our findings in the main text suggest that self-reported preference data are consistent with our hypothesis that highly-central individuals have preferences that are exceptionally similar to each other and less-central individuals have preferences that are idiosyncratic, with each low-centrality individual's preferences differing from those of other individuals in their own way. See Supplementary Tables 7 and 8.

**Supplementary Table 7 and Supplementary Table 8 for "Dyad-level preference analysis" in the "Results" section**

Table S7. Inferring similarities in enjoyment ratings from dyad-level in-degree centralities

| Contrast | $B$ | SE | $p$ (corrected) | $p$ (uncorrected) |
|---|---|---|---|---|
| $s_{\{high, high\}} > s_{\{low, low\}}$ | 0.589 | 0.251 | 0.001** | < 0.001 |
| $s_{\{high, high\}} > s_{\{low, high\}}$ | 0.322 | 0.133 | 0.001** | < 0.001 |
| $s_{\{low, high\}} > s_{\{low, low\}}$ | 0.267 | 0.133 | 0.003** | 0.003 |

* $p < 0.05$, ** $p < 0.01$; we have FDR-corrected all $p$-values because of multiple comparisons; we also report the corresponding uncorrected $p$-values.

Table S8. Inferring similarities in interest ratings from dyad-level in-degree centralities

| Contrast | $B$ | SE | $p$ (corrected) | $p$ (uncorrected) |
|---|---|---|---|---|
| $s_{\{high, high\}} > s_{\{low, low\}}$ | 0.509 | 0.259 | 0.008** | 0.005 |
| $s_{\{high, high\}} > s_{\{low, high\}}$ | 0.277 | 0.137 | 0.008** | 0.004 |
| $s_{\{low, high\}} > s_{\{low, low\}}$ | 0.232 | 0.137 | 0.013* | 0.013 |

* $p < 0.05$, ** $p < 0.01$; we have FDR-corrected all $p$-values because of multiple comparisons; we also report the corresponding uncorrected $p$-values.

**Supplementary Methods 10 for "Dyad-level preference analysis" in the "Results" section**

**Binarized dyad-level ISC results: Controlling for similarities in preference.** We tested whether our findings that individuals who were highly central in their social network tended to have exceptionally similar neural responses to other highly-central individuals and that less-central individuals had idiosyncratic responses could be attributable to inter-subject similarities in self-reported preferences. Although similarities in interest and enjoyment ratings were associated with neural similarity in subregions of the default-mode network (see Supplementary Fig. 11), when we controlled for dyad-level binarized in-degree centrality by including it as a covariate, we found that the associations between similarities in preference ratings and ISCs in many of the brain regions were no longer significant (see Supplementary Fig 12). Notably, we also found that the pattern of results that link ISCs and dyad-level binarized in-degree centrality remain significant after controlling for similarity in enjoyment and interest ratings (see Supplementary Fig. 13). This suggests that our findings that the greater neural similarity occurs between highly-central individuals and that less neural similarity occurs between less-central individuals resulted from differences beyond what we captured using self-reported preference ratings.

**Supplementary Fig. 11 for "Dyad-level preference analysis" in the "Results" section**

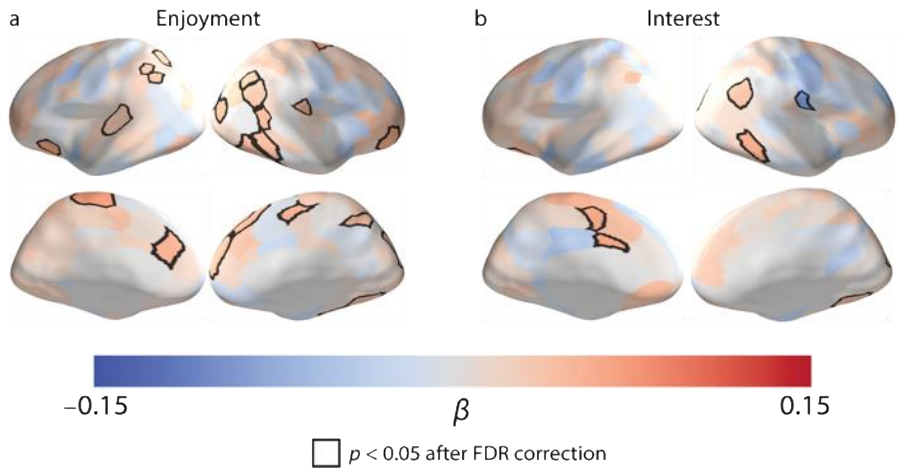

**Supplementary Fig. 11.** Dyad-level associations of neural similarity with similarities in preference ratings. We found that similarities in (a) interest and (b) enjoyment ratings were associated with neural similarity in regions of the default-mode network when not controlling for in-degree centrality. The quantity $B$ is the standardized regression coefficient. Regions where we observed significant associations between similarities in enjoyment (a) and interest (b) ratings and ISC are outlined in black. We used an FDR-corrected significance threshold of $p < 0.05$, which corresponds to an uncorrected $p$-value threshold of $p < 0.006$ for (a) and $p < 0.002$ for (b).

**Supplementary Fig. 12 for "Dyad-level preference analysis" in the "Results" section**

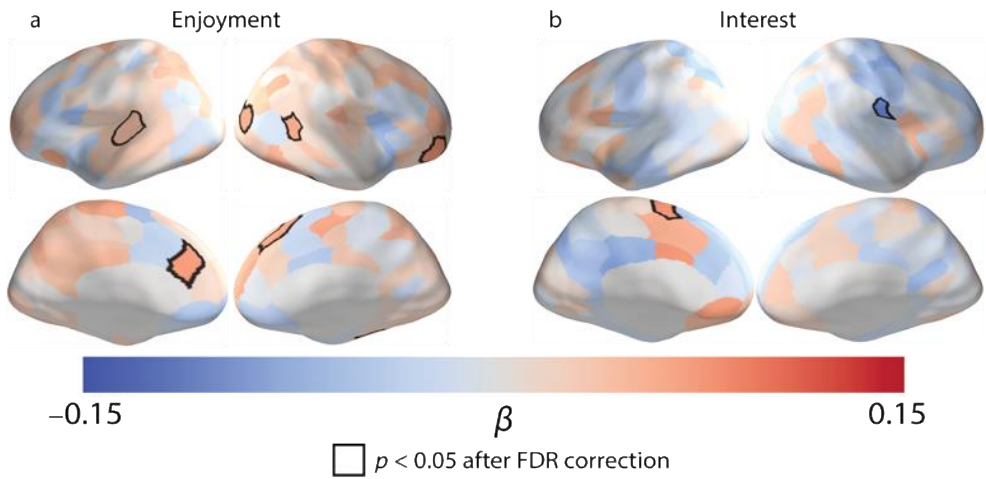

**Supplementary Fig. 12.** Dyad-level associations of neural similarity with similarities in preference ratings when controlling for in-degree centrality. We found that the associations between similarities in preference ratings and ISCs in many of the regions were no longer significant when we controlled for dyad-level binarized in-degree centrality. The quantity $B$ is the standardized regression coefficient. Regions where we observed significant associations between similarities in enjoyment (a) and interest (b) ratings and ISC are outlined in black. We used an FDR-corrected significance threshold of $p < 0.05$, which corresponds to an uncorrected $p$-value threshold of $p < 0.002$ for (a) and $p < 0.001$ for (b).

**Supplementary Fig. 13 for "Dyad-level preference analysis" in the "Results" section**

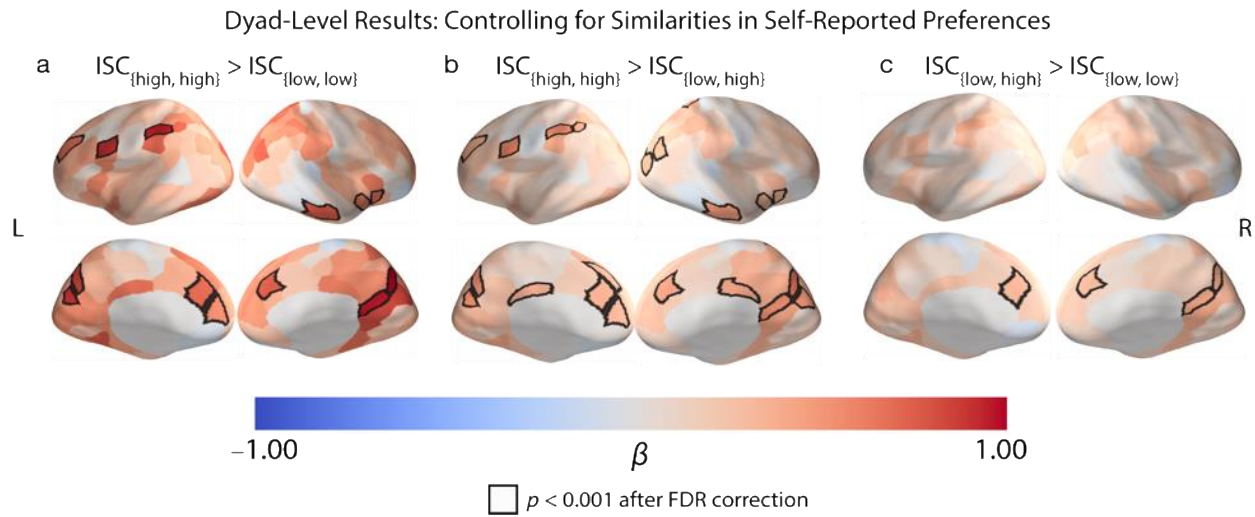

**Supplementary Fig. 13.** Dyad-level associations of neural similarity with in-degree centrality when controlling for similarities in self-reported preferences (specifically, interest and enjoyment ratings) about the videos. We identified similar brain regions as significantly associated with in-degree centrality as in our main manuscript (see Fig. 4). The quantity $B$ is the standardized regression coefficient. Regions where we observed significant associations between in-degree centrality and ISC are outlined in black. We used an FDR-corrected significance threshold of $p < 0.001$, which corresponds to an uncorrected $p$-value threshold of $p < 6.075 \times 10^{-5}$.

## Supplementary Table 9: Table of excluded participants

| Participant | Reason(s) |
| --- | --- |
| Excluded participant 1 | Excessive movement during the fMRI scan |
| Excluded participant 2 | Excessive movement during the fMRI scan |
| Excluded participant 3 | Fell asleep during half of the fMRI scan |
| Excluded participant 4 | Did not complete the fMRI scan and did not complete the social-network survey |
| Excluded participant 5 | Did not complete the social-network survey |
| Excluded participant 6 | Did not complete the social-network survey |
| Excluded participant 7 | Did not complete the social-network survey |